\documentclass[twocolumn, 10pt]{IEEEtran}
\usepackage{cite}
\usepackage{graphicx}
\usepackage{psfrag}
\usepackage{subfigure}
\usepackage{amsmath}
\usepackage{amssymb}
\usepackage{epsfig}
\usepackage{color}
\usepackage{epsf}
\usepackage{algorithmic}
\usepackage{algorithm}
\usepackage{setspace}

\newtheorem{theorem}{Theorem}

\newtheorem{Lemm}{Lemma}
\newtheorem{Result}{Result}

\newtheorem{remark}[theorem]{Remark}

\def\bb0{{\mathbb{0}}}

\def\ba{{\mathbf{a}}}
\def\bb{{\mathbf{b}}}

\def\bff{{\mathbf{f}}}
\def\bg{{\mathbf{g}}}
\def\bh{{\mathbf{h}}}

\def\b0{{\mathbf{0}}}

\def\bA{{\mathbf{A}}}

\def\bG{{\mathbf{G}}}

\def\bI{{\mathbf{I}}}

\def\bbC{{\mathbb{C}}}

\def\bbE{{\mathbb{E}}}

\def\bbP{{\mathbb{P}}}

\def\cK{\mathcal{K}}
\def\cL{\mathcal{L}}

\def\sf0{{\mathsf{0}}}

\def\rmB{\mathrm{B}}

\def\rmE{\mathrm{E}}
\def\rmF{\mathrm{F}}

\def\rmM{\mathrm{M}}

\def\rmb{{\mathrm{b}}}
\def\rmc{{\mathrm{c}}}
\def\rmd{{\mathrm{d}}}

\def\rmm{{\mathrm{m}}}

\def\rms{{\mathrm{s}}}
\def\rmt{{\mathrm{t}}}
\def\rmu{{\mathrm{u}}}

\def\rm0{{\mathrm{0}}}

\def\invSNR{1/{\mathsf{SNR}}}
\def\Nt{{N_\rmt}}

\def\Es{{\rmE_\rms}}

\def\lb{{\lambda_\rmb}}

\begin{document}

\IEEEoverridecommandlockouts

\title{Interference Coordination: Random Clustering and Adaptive Limited Feedback\thanks{S. Akoum and R. Heath are with The University of Texas at Austin, 2501 speedway stop C0806, Austin, TX 78712 USA (e-mail: {salam.akoum, rheath}@mail.utexas.edu). This work was funded in part by Huawei Technologies. }}
\author{
Salam Akoum and Robert W. Heath, Jr.
}
\maketitle

\begin{abstract}
Interference coordination improves data rates and reduces outages in cellular networks. Accurately evaluating the gains of  coordination, however, is contingent upon using a network topology that models realistic cellular deployments.  In this paper, we model the base stations locations as a  Poisson point process to provide a better analytical assessment of the performance of coordination. Since interference coordination is only feasible within clusters of limited size, we consider a random clustering process where cluster stations are located according to a random point process and groups of  base stations associated with the same  cluster coordinate.  We assume channel knowledge is exchanged among coordinating base stations, and we analyze the performance of interference coordination when channel knowledge at the transmitters is either perfect or acquired through limited feedback. We apply intercell interference nulling (ICIN) to coordinate interference inside the clusters. The feasibility of ICIN depends on the number of antennas at the base stations. Using tools from stochastic geometry, we derive the probability of coverage and the average  rate for a typical mobile user.  We show that the average cluster size can be optimized as a function of the number of antennas to maximize the  gains of ICIN. To minimize the mean loss in rate due to limited feedback, we propose an adaptive feedback allocation strategy at the mobile users. We show that adapting the bit allocation as a function of the signals' strength increases the achievable rate with limited feedback,  compared to equal bit partitioning. Finally, we illustrate how this analysis can help solve network design problems such as identifying regions where coordination provides gains based on average cluster size, number of antennas, and number of feedback bits. 
\end{abstract}

\section{Introduction}
Coordination can mitigate interference and increase data rates in cellular systems \cite{Ges2010}. Complete coordination between all the base stations in the network, however, is not feasible \cite{Papadogiannis2008, Lozano2012}. For inter-base station overhead to be affordable, only groups of base stations coordinate, forming coordination clusters \cite{Papadogiannis2008, simeone2009, Zhang2009a}. To quantify the gains from  coordination, an accurate model of the network topology and the relative locations of the mobile users and the base stations needs to be considered \cite{simeone2009}. Much prior analytical work on interference coordination used oversimplified network models and reported coordination gains that did not materialize in practical cellular deployment scenarios \cite{Barbieri2012}. This paper addresses this issue by considering a point process model for deployment and clustering of base stations. 

Most of the literature on interference coordination \cite{Shamai2001, ekbal2005, Jorwiesk2008, Ng2008, simeone2009, Ges2010,  Zhang2009a, Ng2010, Zhang2010, bhagavatula2011, Kaviani2012} considered fixed cellular network architectures such as the Wyner model or the hexagonal grid. The Wyner model is used to derive information theoretic bounds on the performance of multicell cooperation  \cite{Shamai2001, simeone2009, Ges2010}; it does not account for the random locations of the users inside the cells. Hexagonal grid models, although reasonably successful in studying cellular networks, generally rely on extensive Monte Carlo simulations to gain insight into effective system design parameters \cite{Foschini2006, Papadogiannis2008}. 
Using random models for the base stations locations yields, under fairly simple assumptions, analytical characterizations of outage and capacity, and provides a good approximation for the performance of actual base stations deployment\cite{Andrews2010}. 

We consider randomly deployed base stations. To form coordinating base station clusters, we propose a random clustering model, in which cluster stations are randomly deployed in the plane, and base stations connect to their geographically closest cluster station. The random clustering model builds on the analytical appeal of the random network deployment. It mirrors the connection of base transceiver stations to their base station controllers in current cellular systems \cite{CoMP2011}. A similar hierarchical structure was used in \cite{baccelli1997} to model traffic and requests for communication in wired telecommunication networks. A regular lattice clustering model of randomly deployed base stations was considered in \cite{Huang2012j} to derive the asymptotic outage performance of interference coordination as a function of the location of the user inside the fixed grid and the scattering model. 

%
%

To achieve coordination gains, coordinating base stations exchange channel state information (CSI) on the backhaul. Each base station designs its  beamforming vector to transmit exclusively to the users in its own cell, while exchanging CSI of the served users with the cooperating base stations. Examples of interference coordination strategies in the literature \cite{ekbal2005, Jorwiesk2008, Ng2008, Zhang2010, bhagavatula2011} include iterative strategies such as MMSE estimation beamforming \cite{Ng2008} and non-iterative strategies such as intercell interference nulling (ICIN) \cite{Jorwiesk2008, Zhang2010, bhagavatula2011}.  ICIN is a cooperative beamforming strategy where each base station transmits in the null space of its interference channels.

The gains from  coordination depend on the quality of CSI available to the base stations. In frequency division duplex systems,  the CSI of the estimated channels  at the mobile users can be made available at the base stations through feedback. The feedback channel possesses finite bandwidth and thus limited feedback techniques \cite{Love2008} are employed. Most of the literature on multicell coordination \cite{Zhang2009a,Ng2010, Kaviani2012} assumes full CSI at the transmitters, neglecting the performance loss due to quantization. For ICIN with limited feedback, each mobile user quantizes and feeds back the CSI  of the desired and interfering channels. This is in contrast with single-cell limited feedback , where only the CSI of the desired channel is fed back \cite{Akoum2011a}. Dividing the feedback budget among the desired and interfering channels has been proposed as a reasonable approach for ICIN with limited feedback \cite{bhagavatula2011, Ozbek2010}. The allocation of bits to  channels at the receiver can be done equally or adaptively, taking into account the location of the user and the strength of the channel being quantized \cite{bhagavatula2011}. 

In this paper, we analyze the performance gains of interference coordination in a clustered network model where the base station locations are distributed as an independent homogeneous Poisson point process (PPP) and the clustering of the base stations is done through overlaying another independent homogeneous PPP. Base stations belonging to the same cluster coordinate via ICIN. We account for the feasibility of ICIN given the number of interferers in each cluster and the number of antennas at each base station. We distinguish between two cases, first assuming that ICIN is always feasible, and later considering a preset number of antennas at each base station and only applying ICIN when feasible. Single-cell beamforming is applied otherwise. Using tools from stochastic geometry, we derive a bound on the coverage and average rate performance of the proposed clustered coordination model for both cases. For the case of limited feedback CSI, we derive a bound on the mean loss in rate. We assume random vector quantization (RVQ) at the mobile users for analytical reasons. To minimize the mean loss in rate due to quantization, we consider an adaptive bit allocation algorithm and we derive closed form expressions to determine the number of bits to allocate to each channel based on its strength.

The contributions of the paper are summarized as follows.
\begin{itemize}
\item We propose a random hierarchical clustering model based on point process deployment of cluster stations and base stations. We assume that the base stations connect to their geographically closed cluster stations to form coordination clusters.
\item We derive bounds on the coverage and average rate performance of ICIN, for the proposed clustered model, as a function of the number of interferers inside each cluster, the number of antennas at each base station, and the average cluster size.
\item We analyze  the feasibility of ICIN taking into account the random number of interferers inside each cluster and the number of antennas at the base stations. We propose a thresholding policy where ICIN is applied only when feasible, and single-cell beamforming is applied otherwise. We derive bounds on the probability of coverage and average rate performance of the thresholding policy. We show that there is an optimal cluster size to achieve maximum coordination gains.
\item We derive an upper bound on the mean loss in rate due to limited feedback using ICIN. The bound is a function of the number of feedback bits allocated to each channel, the number of interferers, and the relative channel strengths at the mobile user. We derive closed form expressions for allocating the feedback bits to the desired and interfering channels, as a function of their relative strengths at the mobile user, and taking into account the relative strength of the inter-cluster and intra-cluster interference. 
\end{itemize}

Throughout the paper, we use the following notation, $\ba$ is a column vector, $\bA$ is a matrix, $a$ is a scalar. $\bA^*$ denotes the conjugate transpose and $\bA^{-1}$ the matrix inverse. The left pseudo-inverse of $\bA$ with linearly independent columns is $\bA^\dagger =  \left(\bA^*\bA\right)^{-1}\bA^*$. $\bbE\{.\}$ denotes expectation. $|.|$ denotes the cardinality of a set.


 \begin{figure}[t]
  \begin{center}
  \vspace{-10pt}
    \includegraphics[scale = 0.35]{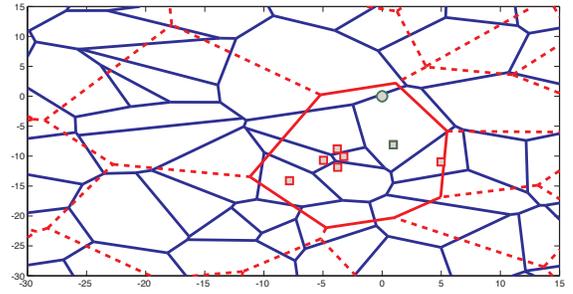}
    \caption{The hierarchical cellular model considered. The Voronoi tessellation of the plane formed by the cells of $\Pi_\rmb$ is shown in blue. The Voronoi tessellation formed by $\Pi_\rmc$ is shown in red. The typical mobile user is shown in a green circle.  The cooperating base stations for the cluster of interest are shown in rectangles.}
    \label{fig:hierarchicalModel}
  \end{center}
\end{figure}
\section{System Description and Assumptions}\label{sec:sysModel}
Consider the network model shown in Figure~\ref{fig:hierarchicalModel}. The base stations are represented by an independent homogeneous PPP $\Pi_\rmb$ of density $\lambda_\rmb$. The mobile users are located according to an independent point process $\Pi_\rmu$. The cells of the base stations form a Voronoi tessellation of the plane with respect to the process $\Pi_\rmb$, and the users are connected to their closest base station. We assume that the density of the mobile process is sufficiently large such that all the base stations are active. Each base station is equipped with $\Nt$ antennas and serves one single antenna mobile user in each cell using intra-cell time division multiple access (TDMA). While TDMA is not necessarily the best option for a multiple-input-single-output (MISO) transmission strategy, it is  a common assumption made in the multicell cooperation literature due to the  tractability of single-user transmission \cite{Ges2010},\cite{bhagavatula2011},\cite{simeone2009},\cite{Huang2012j}.

To model clustering, we overlay another independent homogeneous PPP $\Pi_\rmc$ on the 2-D plane with intensity $\lambda_\rmc \leq \lambda_\rmb$. We denote the cluster centers of this PPP  by $\rmc_k$. The cluster cells  form a Voronoi tessellation of the plane with respect to the process $\Pi_\rmc$. The base stations of the $\Pi_\rmb$ process located in the same Voronoi cell $V^{(\rmc_k)}$ of a cluster center $\rmc_k$ coordinate via ICIN. The association of the base stations to the cluster base stations follows similar to the association of the mobile users to their closest base station in the basic cellular model. The cluster base stations can be seen as central processors to which base stations in the same cluster are connected via backhaul. The backhaul is assumed error and delay free. 

Based on the stationarity of the Poisson process \cite{Baccelli2009a}, we consider the performance for a typical user $\rmu_0$ served by a base station $\rmb_0$  inside a  cluster  $V^{(\rmc_0)}(\rmb_0)$.  The channel corresponding to the desired signal between $\rmb_0$ and $\rmu_0$ is denoted by $\bh^*_0 \in \bbC^{1\times \Nt}$. The interfering channels from the $\ell$-th base station to $\rmu_0$ are denoted by $\bg^*_{0,\ell}\in \bbC^{1\times \Nt}$. The desired and interfering channels are modeled according to the Rayleigh fading model, where each entry is  independently and identically distributed as a unit variance zero mean complex Gaussian. The symbol transmitted from the $k$-th base station is denoted by $s_k$, where $\bbE[|s_k|^2] = \rmE_\rms$ and there is no power control. While power control is an important topic for practical networks, an exhaustive investigation of the topic would require a detailed analysis of a distributed power control algorithm among the base stations. Another way of implementing power control independently at each base station is to perform channel inversion and assume fixed received power \cite{Weber2011}. Such a strategy leads to better outage performance for full CSI \cite{Huang2012j}. In the absence of interference coordination, each user is subject to interference from all other  base stations in $\Pi_\rmb$, each transmitting with energy $\rmE_\rms$. The path-loss incurred by the desired signal is given by $L(r_0) = (1+r_0)^{\alpha}$ where $r_0$ is the distance between $\rmu_0$ and $\rmb_0$, and $\alpha$ is the path-loss exponent. For the interfering signals from the $\ell$-th base station to $\rmu_0$, the path-loss is given by $L(r_{0,\ell}) = (1+ r_{0,\ell})^{\alpha}, r_{0,\ell} >r_0$. Most of the literature on analysis of random spatial models  \cite{Andrews2010, Huang2012j} consider the $L(r) = r^{\alpha}$ path-loss model; this model, however, has a singularity at zero and is inaccurate for small distances. The received signal powers of the desired and interfering signals are then given by $\gamma_0 = \rmE_\rms/L(r_0)$ and $\gamma_{0,\ell} = \rmE_\rms/L(r_{0,\ell})$. Using a narrowband flat-fading model, the baseband discrete-time input-output relation for $\rmu_0$ is given by
\begin{equation}
y_0 = \sqrt{\gamma_0}\bh^*_0\bff_0s_0 + \sum_{\rmb_\ell \in \Pi_{\rmb}/\rmb_0}{\sqrt{\gamma_{0,\ell}}\bg^*_{0,\ell}\bff_\ell s_\ell} + v_0,
\end{equation}
where $y_0$ is the received signal at $\rmu_0$, the vector $\bff_0 \in \bbC^{\Nt\times 1}$ is the unit-norm beamforming vector at $b_0$, and $\bff_\ell \in \bbC^{\Nt\times 1}$ is the unit-norm beamforming vector at the $\ell$-th base station. The scalar $v_0$ denotes the additive white Gaussian noise at the receiver with variance $\sigma^2$. 

The signal-to-interference-plus-noise ratio (SINR) at $\rmu_0$ is given by
\begin{eqnarray}
\mathsf{SINR}_0 &=& \frac{\gamma_0|\bh^*_0\bff_0|^2}{\sigma^2 + \sum_{\rmb_\ell \in \Pi_{\rmb}/\rmb_0}{\gamma_{0,\ell}|\bg^*_{0,\ell}\bff_\ell|^2}}.
\end{eqnarray}
For a given SINR threshold $T$, a performance metric of interest is the probability of coverage,
\begin{eqnarray}
p_c(\lambda_\rmb, \lambda_\rmc,\alpha,T) = \bbP\left[ \mathsf{SINR}_0 \geq T \right].
\end{eqnarray}
The probability of coverage is the probability that a randomly chosen user in the 2-D plane has a target SINR greater than $T$. Such a condition is required in practice when a given constant bit-rate corresponding to a particular coding scheme needs to be sustained.

Another performance metric of interest is the average  rate,
\begin{eqnarray}
\tau\left(\lb, \lambda_\rmc,\alpha \right) =  \bbE\left\{\log_2\left(1 + \mathsf{SINR}_0\right) \right\}.
\end{eqnarray}
 To express the average  rate, we assume that adaptive modulation/coding is used so that the users achieve a Shannon bound for their rate. 
 
\section{Per Cluster Interference Coordination}\label{sec:intfCoord}
We propose interference coordination within each cluster based on zero forcing (ZF) intercell interference nulling. We assume there are $N$ interfering base stations in the cluster of interest. Since $N$ is a random variable that depends on $\lambda_\rmc$ and $\lambda_\rmb$, the number of antennas at the base stations needs to be sufficiently high ($\Nt > N)$ for ICIN to be feasible inside each cluster. We assume for the first part of this paper that $\Nt$ grows with $N$ such that $\Nt - N = \rmd_\Nt$, where $\rmd_\Nt$ are the extra degrees of freedom at each base station, used to beamform to the desired user \cite{Jindal2011, Akoum2011}. This assumption is relaxed to a preset constant $\Nt$ in Section \ref{sec:thresh}. Each base station in the cluster has knowledge not only of the channel to its intended receiver, but also of the interference  channels towards the receivers in the same cluster. Without loss of generality, for the case of the typical user, $\rmu_0$ estimates its desired channel $\bh_0$ and the interfering channels $\bg_{0,\ell}, \ell = 1, \cdots, N$, and feeds the information back to its base station $\rmb_0$. The base stations in the coordination cluster then exchange the interfering CSI on the backhaul, so that $\rmb_0$ has knowledge of $\bh_0$ and the interference channels towards other receivers in the cluster $\bg_{\ell,0}, \ell = 1,\cdots, N$, 

We assume that perfect CSI is available at the receivers. Accounting for imperfections due to channels downlink training further degrades the achievable rate with imperfect CSI, and is a subject of future investigation. We consider first the ideal case where full CSI is available at the transmitters, we then discuss finite rate feedback.
\subsection{ICIN with Perfect CSI}
With perfect CSI at the transmitters, each base station designs its beamforming vector such that it is the normalized projection of the desired channel direction onto the null space of the matrix of interfering channel directions. For $\rmb_0$,  $\bff_0$ is the normalized projection of $\widetilde{\bh}_0= \bh_{0}/\|\bh_{0}\|$ onto the null space of  $\bG_{I,0} = \left[\widetilde{\bg}_{1,0}\cdots \widetilde{\bg}_{\ell,0}\cdots \widetilde{\bg}_{N,0} \right]$, with $\widetilde{\bg}_{\ell,0} = {\bg}_{\ell,0}/\|\bg_{\ell,0}\|$, 

\small
\begin{eqnarray}
\bff_0 = \frac{\left(\bI - \bG_{I,0}\bG_{I,0}^{\dagger}\right)\widetilde{\bh}_0}{\left\|\left(\bI - \bG_{I,0}\bG_{I,0}^{\dagger}\right)\widetilde{\bh}_0\right\|}.
\end{eqnarray}
\normalsize
The  $\mathsf{SINR}_0$ with per cluster ICIN, is given by

\small\begin{eqnarray}
\nonumber\mathsf{SINR}_{\mathrm{ic}} &=&  \frac{\gamma_0|\bh^*_0\bff_0|^2}{\sum_{\rmb_\ell \in  \Pi_{\rmb}/V^{(\rmc_0)}(\rmb_0)}{\gamma_{0,\ell}|\bg^*_{0,\ell}\bff_\ell|^2} + \sigma^2} \\
&=&  \frac{(1+ r_0)^{-\alpha}|\bh^*_0\bff_0|^2}{I_{\mathrm{out}}+\invSNR},
\end{eqnarray}\normalsize
where $I_{\mathrm{out}}$ is the inter-cluster interference caused by all the base stations outside the cluster of interest $V^{(\rmc_0)}(\rmb_0)$, denoted by $\Pi_{\rmb}/V^{(\rmc_0)}(\rmb_0)$. The signal-to-noise-ratio (SNR) here is defined in terms of the transmit signal power, $\invSNR = {\sigma^2}/{\Es}$.

\subsection{ICIN with Limited Feedback CSI}
With limited feedback, the estimated channel directions $\widetilde{\bh}_0$ and $\widetilde{\bg}_{0,\ell}$ are quantized to the unit-norm vectors given by $\widehat{\bh}_0$ and $\widehat{\bg}_{0,\ell}$, respectively, at $\rmu_0$. These quantized channel directions are then fed back to the base stations using a fixed feedback budget of $\rmB_{\mathrm{tot}}$ bits. The base stations use these quantized channel directions to design the beamforming vector. We assume that the base stations have perfect knowledge of the SNR at the receivers, independently of the channel directions. Perfect non-quantized knowledge of the received SNR at the base stations is a common assumption in the literature on multi-user and multicell limited feedback \cite{Sharif2005, bhagavatula2011}.

We consider separate quantization at the receiver of the desired and interfering signals. Joint quantization was considered in \cite{bhagavatula2010}; it requires a large storage space at the mobile users.
Each user divides $\rmB_{\mathrm{tot}}$ among the desired and interfering channels such that $\rmB_0$ and $\rmB_{0,\ell}$ are used to quantize $\widetilde{\bh}_0$ and $\widetilde{\bg}_{0,\ell}$ respectively, and $\rmB_0 + \sum_{\ell=1}^{N}{\rmB_{0,\ell}} = \rmB_{\mathrm{tot}}$ bits. We ignore delays on the feedback channel as well as delays on the backhaul link between the base stations belonging to the same cluster. The beamforming vector $\widehat{\bff}_0$ is the normalized projection of the quantized channel $\widehat{\bh}_0$ onto the null space of the matrix $\widehat{\bG}_{I,0} = \left[\widehat{\bg}_{1,0} \cdots \widehat{\bg}_{\ell,0}\cdots \widehat{\bg}_{N,0} \right]$,

\small\begin{eqnarray}
\widehat{\bff}_0 = \frac{\left(\bI - \widehat{\bG}_{I,0}\widehat{\bG}_{I,0}^{\dagger}\right)\widehat{\bh}_0}{\left\|\left(\bI - \widehat{\bG}_{I,0}\widehat{\bG}_{I,0}^{\dagger}\right)\widehat{\bh}_0\right\|}.
\end{eqnarray}
\normalsize
$\mathsf{SINR}_0$ with limited feedback and per cluster ICIN, is given by

\small\begin{eqnarray}
\nonumber \widehat{\mathsf{SINR}}_{\mathrm{ic}}&=& 
 \frac{(1+r_0)^{-\alpha}|\bh^*_0\widehat{\bff}_0|^2}{\frac{1}{\mathsf{SNR}}  ~+~\widehat{I}_{\mathrm{out}}~+~I_{\mathrm{res}}},
\end{eqnarray}\normalsize
where $I_{\mathrm{res}}=\sum_{\rmb_\ell \in V^{(\rmc_0)}(\rmb_0)/\rmb_0}{(1+r_{0,\ell})^{-\alpha}|\bg^*_{0,\ell}\widehat{\bff}_\ell|^2} $ is the residual intra-cluster interference due to quantization. Interference signals are not nulled out, since $\bg^*_{0,\ell}\widehat{\bff}_\ell \neq 0$ but  $\widehat{\bg}^*_{0,\ell}\widehat{\bff}_\ell = 0$. $\widehat{I}_{\mathrm{out}} =\sum_{\rmb_\ell \in \Pi_\rmb/V^{(\rmc_0)}(\rmb_0)}{(1+r_{0,\ell})^{-\alpha}|\bg^*_{0,\ell}\widehat{\bff}_\ell|^2}$ is the inter-cluster interference assuming limited feedback beamforming. 

We derive the average  rate with perfect CSI in Section \ref{sec:perfEval}. We then bound the mean loss in rate with limited feedback for equal bit and adaptive bit partitioning among the desired and interfering channels  in Section \ref{sec:impLim}.

\section{Performance Evaluation with Perfect CSI}\label{sec:perfEval}
In this section, we analyze the outage performance of per cluster ICIN. We then use the probability of coverage expression to derive a bound on  the average  rate.

\subsection{Probability of Coverage Analysis}
The probability of coverage is the complementary cumulative distribution (CCDF) of $\mathsf{SINR}_{\mathrm{ic}}$,

\small\begin{eqnarray}\label{eqn:pcoverage}
 p_c(\lambda_\rmb,\lambda_\rmc, \alpha,T) &=& \bbP\left[ \mathsf{SINR}_{\mathrm{ic}} \geq T \right]\\
\nonumber &=& \bbP\left[{|\bh^*_0\bff_0|^2}\geq {T L(r_0)\left(I_{\mathrm{out}}+\invSNR\right)}\right].
\end{eqnarray}\normalsize
The desired effective channel power $|\bh^*_0\bff_0|^2$ is Gamma-distributed with parameters $\rmd_\Nt$ and $1$, $|\bh^*_0\bff_0|^2 \sim \Gamma\left[\rmd_\Nt, 1\right]$. This follows from the projection of the independent normal Gaussian isotropic vector $\bh_0$ onto the null space of the interfering channels of dimension $\rmd_\Nt = \Nt - N$ \cite{Jindal2011, Akoum2011a}. 

The aggregate inter-cluster interference power $I_{\mathrm{out}}$ is a function of the cluster size. While a numerical fit of the distribution of the cluster size is possible \cite{ferenc2007}, an expression for the exact size of the Voronoi region $V^{(\rmc_0)}(\rmb_0)$ is hard to compute. Hence, we bound the area of the cluster of interest by the area of the maximal disk $\rmB_\rmm$ inscribed in the cluster, and centered at $\rmc_0$. Let $r_\rmm$ designate the radius of $\rmB_\rmm$. $r_\rmm$ is Rayleigh distributed with CCDF 
$\bbP[r_m>r] = \exp(-4\pi\lambda_\rmc r^2)$, \cite{Foss96}.
The interference power is then upper bounded by

\small\begin{eqnarray}\label{eqn:IoutIneq}
\nonumber I_{\mathrm{out}} &=&  \sum_{\Pi_{\rmb}/V^{(c)}(\rmb_0)}{\gamma_{0,\ell}|\bg^*_{0,\ell}\bff_\ell|^2} \leq \sum_{\Pi_{\rmb}/\rmB_\rmm}{\gamma_{0,\ell}|\bg^*_{0,\ell}\bff_\ell|^2}\\
&\leq& \sum_{\Pi_{\rmb}/\rmB_\rmm(\rmu_0)}{\gamma_{0,\ell}|\bg^*_{0,\ell}\bff_\ell|^2},
\end{eqnarray}\normalsize
where $\rmB_\rmm(\rmu_0)$ denotes the disk centered at the typical mobile user $\rmu_0$ of radius $r_\rmm - r_0 - r_1$, where $r_1$ is the distance from $\rmb_0$ to $\rmc_0$, Rayleigh distributed with probability density function (PDF) $f_{r_1}(r) = 2\pi \lambda_{\rmc} r \exp({-\pi\lambda_{\rmc} r^2})$. The interference field power is computed at $\mathrm{u}_0$ located at a distance $0 \leq r \leq r_\mathrm{m}$ from the center of the inscribed disk $\mathrm{B}_\mathrm{m}$ corresponding to the cluster-cell center. The exclusion distance to the nearest interferer from the mobile user is asymmetric since the distance from the mobile user to the closest edge of the disk is smaller than its distance from the furthest edge. To avoid the dependence on the location of the interference field, we pursue an upper bound on the interference power. We consider the interference contribution in a disk of radius $r_\mathrm{m} - r_1 - r_0$ centered at $\mathrm{u}_0$ and denoted by $\mathrm{B}_\mathrm{m}(\mathrm{u}_0)$.  The exclusion areas are such that $\mathrm{B}_\mathrm{m}(\mathrm{u}_0) \subseteq \mathrm{B}_\mathrm{m} \subseteq V^{(\mathrm{c}_0)}(\mathrm{b}_0)$.
The upper bound on  the aggregate inter-cluster interference power results in a lower bound on the probability of coverage.
\begin{Result}\label{lem:pic}
The probability of coverage with per cluster interference coordination is lower bounded by

\footnotesize\begin{align}\label{eqn:p_ic}
\nonumber &p_{\rmc,\mathrm{ic}}(\lambda_\rmb,\lambda_\rmc, \alpha, T)\geq \left\{\int_{0}^{\infty}{f_r(r_0)\displaystyle\int_{r_0}^{\infty}f_{r_\rmm}(r_\rmm)}\times\right.\\
 &\left.{{\int_{-\infty}^{\infty}{e^{-2\pi j\frac{TL(r_0)}{\mathsf{SNR}} s}\cL_{I_{r_\rmm }}\left(2j\pi L(r_0)T s\right)\frac{\cL_h(-2j\pi s)-1}{2i\pi s}\rmd s}\rmd r_0} \rmd r_\rmm}\right\},
\end{align}\normalsize
where the Laplace transform of the desired signal $|\bh^*_0\bff_0|^2 \sim \Gamma[\rmd_\Nt,1]$, is

\small\begin{eqnarray}
\nonumber \cL_h(s) = \frac{1}{\left(s + 1\right)^{\rmd_\Nt}}, \quad {\rmd_\Nt = \Nt - N}.
\end{eqnarray}\normalsize
The PDF of $r_0$, $f_{r}(r) = 2\pi \lambda_{\rmb} r \exp({-\pi\lambda_{\rmb} r^2})$, follows from the null probability of a 2-D PPP with respect to $\Pi_{\rmb}$.  The PDF of the radius $r_\rmm$ is  
$f_{r_\rmm}(r) = 8\pi \lambda_{\rmc} r \exp({-4\pi\lambda_{\rmc} r^2})$.
The Laplace transform of the interference $\cL_{I_{r}}$ is

\footnotesize\begin{align}
 &\cL_{I_{r}}(s) =\\
\nonumber& \exp\left\{-2\pi\lambda_\rmb\left(\frac{(1+r)^{2-\alpha}}{\alpha-2}s \;{}_2\rmF_1\left(1,1-\frac{2}{\alpha}, 2-\frac{2}{\alpha}, -(1+r)^{-\alpha} s\right) \right.\right.\\
\nonumber &- \left.\left.\frac{(1 + r)^{1-\alpha}}{\alpha-1} s\;{}_2\rmF_1\left(1,1 -\frac{1}{\alpha},2 - \frac{1}{\alpha}, -{(1+r)^{-\alpha}}{s} \right) \right)\right\},
\end{align}\normalsize
where ${}_2\rmF_1(a,b,c,z) = \frac{\Gamma(c)}{\Gamma(b)\Gamma(c-b)}\int_{0}^{1}{\frac{t^{b-1}(1-t)^{c-b-1}}{(1-tz)^a}}$ denotes the Gauss hypergeometric function.
\end{Result}
\begin{IEEEproof}
See Appendix \ref{app:app_1}.
\end{IEEEproof}
The tightness of the upper bound on the probability of outage is investigated in Section \ref{sec:simRes}. The probability of coverage depends on the density of interfering base stations, the density of the clusters, the path-loss exponent and the number of extra spatial dimensions available at the transmitters $\rmd_\Nt$.
It is an increasing function of the ratio of the density of interfering base stations to the density of the cluster base stations $\lambda_\rmb/\lambda_\rmc$, which is also the average number of base stations per cluster. It is also an increasing function of $\rmd_\Nt$. As $\Nt$ increases, $\rmd_\Nt$ increases, and the signal power at the mobile users increases, which incurs an increase in the SINR, in addition to the decrease in the interference term due to ICIN per cluster.

\subsection{Average  Rate}
The average rate follows from the probability of coverage analysis as follows

\footnotesize\begin{eqnarray}\label{eqn:IEEEproof_th}
\tau &=& \bbE\left\{\log_2\left(1 + \mathsf{SINR}_0\right) \right\}= \int_{0}^{\infty}{\frac{\bbP\left[\mathsf{SINR}_0 > e^t-1\right]}{\log(2)}dt}.
\end{eqnarray}\normalsize
Inserting  (\ref{eqn:p_ic}) into (\ref{eqn:IEEEproof_th}) gives a lower bound on the average  rate of the typical user.
\begin{Result}
The throughput of the typical user $\rmu_0$, averaged over the spatial realizations of the point processes $\Pi_\rmb$ and $\Pi_\rmc$, and the channel fading distributions is bounded by

\footnotesize\begin{eqnarray}
 \nonumber \tau_{\mathrm{ic}}(\lambda_{\rmb},\lambda_{\rmc},\alpha)  &=& \int_{0}^{\infty}{\frac{\bbP\left[\mathsf{SINR}_\mathrm{ic} > e^t-1\right]}{\log(2)}dt}\\
 \nonumber &=&  \int_{0}^{\infty}{\frac{p_{\rmc,\mathrm{ic}}\left(\lambda_\rmb,\lambda_\rmc,\alpha,v\right)}{(1+v)\log(2)}dv}\\
 &\geq&  \int_{0}^{\infty}{\frac{p^{\mathrm{lb}}_{\rmc,\mathrm{ic}}\left(\lambda_\rmb,\lambda_\rmc,\alpha,v\right)}{(1+v)\log(2)}dv},
\end{eqnarray}\normalsize
where $p^{\mathrm{lb}}_{\rmc,\mathrm{ic}}\left(\lambda_\rmb,\lambda_\rmc,\alpha,v\right)$ is the lower bound on the probability of coverage  $p_{\rmc,\mathrm{ic}}\left(\lambda_\rmb,\lambda_\rmc,\alpha,v\right)$ on the right hand side of (\ref{eqn:p_ic}).
\end{Result}

The computation of $\tau_{\mathrm{ic}}$ requires an additional numerical integration over $p_{\rmc,\mathrm{ic}}$. As in the case of coverage, $\tau_{\mathrm{ic}}$ depends on the average cluster size. It increases with increasing $\rmd_\Nt$ and increasing $\lambda_\rmb/\lambda_\rmc$, as illustrated in Section \ref{sec:simRes}.

\section{Impact of Limited Feedback on Average Rate}\label{sec:impLim}
The performance of ICIN depends on the availability of accurate channel state information at the cooperating base stations. With limited feedback, the quantized CSI is fed back from the receiver to the transmitter. The feedback resources at each mobile user are partitioned among the desired and interfering channels. The bit allocation strategy between the quantized channels is hence important to minimize the mean loss in rate due to limited feedback. 

We compute an upper bound on the mean loss in rate using limited feedback. We use separate codebooks to quantize each desired and interfering channel. We consider random vector quantization in which each of the quantization vectors is independently chosen from the isotropic distribution on the $\Nt$ dimensional unit sphere \cite{Jindal2006}. RVQ is chosen because it is amenable to analysis and yields tractable bounds on the mean loss in rate. Furthermore, optimum codebooks are not yet known for multicell cooperative transmission \cite{bhagavatula2011}.

We define the mean loss in rate due to limited feedback as 

\footnotesize\begin{align}\label{eqn:totalloss}
 {\Delta\widehat{\tau}_{\mathrm{ic}}} = \bbE\bigg\{\log_2\left(1 + \mathsf{SINR}_{\mathrm{ic}}\right)\bigg\} - \bbE\bigg\{\log_2\left(1 + \widehat{\mathsf{SINR}}_{\mathrm{ic}} \right)\bigg\} = {\tau}_{\mathrm{ic}}  - \widehat{\tau}_{\mathrm{ic}},
\end{align}\normalsize
where $\widehat{\tau}_{\mathrm{ic}}$ is the average  rate with limited feedback\footnote{$\bbE\bigg\{\log_2\left(1 + \widehat{\mathsf{SINR}}_{\mathrm{ic}} \right)\bigg\}$ is an upper bound on the achievable rate with limited feedback \cite{Caire2010a}. This bound is approached when the mobile user has perfect knowledge of the coupling coefficients between the beamforming vectors used at the base stations and the channel vectors \cite{Caire2010a}.}.


The SINRs with perfect and limited feedback CSI are such that \small{$\mathsf{SINR}_{\mathrm{ic}} \geq \widehat{\mathsf{SINR}}_{\mathrm{ic}}$}\normalsize. Subsequently \small{$\log_2(1+ \mathsf{SINR}_{\mathrm{ic}})  - \log_2(1+ \widehat{\mathsf{SINR}}_{\mathrm{ic}}) \leq  \log_2( \mathsf{SINR}_{\mathrm{ic}})  - \log_2(\widehat{\mathsf{SINR}}_{\mathrm{ic}})$} \normalsize and 
\vspace{-0.05in}

\small\begin{eqnarray}
\Delta\widehat{\tau}_{\mathrm{ic}} \leq  \bbE\{\log_2( \mathsf{SINR}_{\mathrm{ic}})  - \log_2(\widehat{\mathsf{SINR}}_{\mathrm{ic}})\}. 
\end{eqnarray}\normalsize
We denote the upper bound on $\Delta\widehat{\tau}_{\mathrm{ic}}$ by $\Delta\widehat{\tau}^{\mathrm{ub}}_{\mathrm{ic}}$ given by

\footnotesize\begin{eqnarray}\label{eqn:deltaTau_ic}
\nonumber\Delta\widehat{\tau}^{\mathrm{ub}}_{\mathrm{ic}} = \underbrace{\bbE\bigg\{\log_2\left(\frac{|\widetilde{\bh}^*_0\bff_0|^2}{|\widetilde{\bh}^*_0\widehat{\bff}_0|^2}\right)\bigg\}}_{\Delta\widehat{\tau}_{\mathrm{des}}}
  +  \underbrace{\bbE\bigg\{\log_2\left(\frac{\widehat{I}_{\mathrm{out}}+\frac{1}{\mathsf{SNR}}+ I_{\mathrm{res}}}{I_{\mathrm{out}}+\frac{1}{\mathsf{SNR}}}\right) \bigg\}}_{\Delta\widehat{\tau}_{\mathrm{int}}},
\end{eqnarray}\normalsize
where $\Delta\widehat{\tau}_{\mathrm{des}}$ denotes the mean loss from quantizing the desired channel and $\Delta\widehat{\tau}_{\mathrm{int}}$ is the mean loss from quantizing the interference channels.

To derive the contribution of the desired signal quantization, $\Delta\widehat{\tau}_{\mathrm{des}}$, we use a lower bound on the desired signal power with limited feedback, \cite{bhagavatula2011}

\footnotesize\begin{eqnarray}
 \nonumber \bbE\left\{\log_2\left(|\bh^*_0\widehat{\bff}_0|^2\right)\right\} &\geq& -\bbE_N\bigg\{\frac{\log_2(e)}{\Nt-1}\sum_{i=1}^{\Nt-1}{\beta\left(2^{\rmB_0},\frac{i}{\Nt-1}\right)}\bigg\}\\
 &+& \bbE\left[\log_2\left(\|\bh_0\|^2|\widehat{\bh}_0^*\widehat{\bff}_0|^2\right)\right].
\end{eqnarray}\normalsize
An upper bound on $\Delta\widehat{\tau}_{\mathrm{des}}$ then follows 

\footnotesize\begin{eqnarray}
\Delta{\widehat{\tau}_{\mathrm{des}}}\leq \log_2(e)\;\bbE_{N}\bigg\{ \frac{1}{\Nt-1}\sum_{i=1}^{\Nt-1}{\beta\left( 2^{\rmB_{0}},\frac{i}{\Nt-1}\right)}\bigg\}
\end{eqnarray}\normalsize
since  \small{$\bbE\bigg\{\log_2\left(\|\bh_0\|^2|\widehat{\bh}_0^*\widehat{\bff}_0|^2\right)\bigg\} = \bbE\bigg\{\log_2\left(\|\bh_0\|^2|{\bh}_0^*{\bff}_0|^2\right)\bigg\}$}\normalsize.  The Beta function is defined as \small$\beta(a,b) = \int_0^{1}{t^{a-1}(1-t)^{b-1}dt}$\normalsize.

The effective interference power from each base station outside the cluster is a function of the interference channels towards the typical user, and the quantization vector $\widehat{\bff}_\ell$. The quantization vector is independent of $\bg^*_{0,\ell}$ and is computed using RVQ. The interference power from each base station is thus distributed as  $|\bg^*_{0,\ell}\widehat{\bff}_\ell|^2 \sim \Gamma[1,1], \ell \in {1,\cdots, N}$, \cite{Jindal2006, Weber2011}. We rewrite $\Delta\widehat{\tau}_{\mathrm{int}}$ as

\footnotesize\begin{eqnarray}\label{eqn:deltaTint}
\nonumber \Delta\widehat{\tau}_{\mathrm{int}} =   \bbE\bigg\{\log_2\left({{I}_{\mathrm{out}}+\frac{1}{\mathsf{SNR}}+ I_{\mathrm{res}}}\right) - \log_2\left({I_{\mathrm{out}}+\frac{1}{\mathsf{SNR}}}\right) \bigg\},
\end{eqnarray}\normalsize
and we focus hereafter on the effect of the residual intra-cluster interference $I_{\mathrm{res}}$. 

The statistics of $I_{\mathrm{res}}$ and  the bound on the mean loss in rate depends on the number of interfering base stations inside the cluster $N$, and the strategy of allocating bits among the desired and interfering channels. We first consider an equal bit allocation strategy, independent of the channels' strength, we then optimize the bit allocation to minimize the mean loss in rate.

\subsection{Equal Bit Allocation}
One option for bit allocation is to divide $\rmB_{\mathrm{tot}}$ almost equally between the interfering channels and the desired channel, that is $\rmB_{0,\ell} = \lfloor \rmB_{\mathrm{tot}}/(N+1)\rfloor$ and then set $\rmB_{0} = \rmB_{\mathrm{tot}} - N\rmB_{0,\ell}$. This is a suboptimal strategy because it does not account for the difference in the path-loss between the various interferers, and hence the difference in the importance of the interfering signals. It gives, however, a slight bias for quantizing the desired channel, when the number of bits is not a multiple of the number of base stations in the cluster.  In this section, we compute a closed form expression for the bound on the mean loss in rate, and we use it as a stepping stone to later optimize the bit allocation. 

The bound on the mean loss in rate is rewritten as
\vspace{-0.1in}

\footnotesize\begin{eqnarray}
\nonumber  \Delta{\widehat{\tau}^{\mathrm{ub}}}_{\mathrm{ic}}&\leq& \log_2(e)\;\bbE_{N}\bigg\{ \frac{1}{\Nt-1}\sum_{i=1}^{\Nt-1}{\beta\left( 2^{\rmB_{0}},\frac{i}{\Nt-1}\right)}\bigg\}\\
\nonumber &-& \bbE\bigg\{\log_2\left({I_{\mathrm{out}}+1/{\mathsf{SNR}}}\right) \bigg\}\\
 \nonumber &+&  \bbE\bigg\{ \log_2\left(\sum_{\ell = 1}^{N}{{(1+r_{0,\ell})^{-\alpha}}|\bg^*_{0,\ell}\widehat{\bff}_\ell|^2} + I_{\mathrm{out}}+1/{\mathsf{SNR}}\right)\bigg\}.
\end{eqnarray}\normalsize
As $\Delta{\widehat{\tau}}^{\mathrm{ub}}_{\mathrm{ic}}$ is averaged over all the spatial realizations in the network, with different number of interferers $N$, we  first derive the probability mass function (PMF) of  $N$ in a typical cluster.  The PMF of $N$ depends on the cluster size. It is the distribution of the number of points inside a  cluster area, given that a randomly chosen point is located in that area. 
\begin{Lemm}\label{rem:pN}
The probability mass function of the number of interferers $N$ inside the cluster of interest is given by

\small\begin{eqnarray}\label{eqn:p_N}
\bbP\left[N = n\right] = \frac{3.5^{4.5}\Gamma(n+4.5)(\lambda_\rmb/\lambda_\rmc)^n}{\Gamma(4.5)n!(\lambda_\rmb/\lambda_\rmc + 3.5)^{n+4.5}},
\end{eqnarray}\normalsize
\end{Lemm}
\begin{IEEEproof}
See Appendix \ref{app:app_3}.
\end{IEEEproof}

For the contribution of the inter-cluster interference $I_\mathrm{out}$ and consequently $I_{r_\rmm}$ in  $ \Delta{\widehat{\tau}^{\mathrm{ub}}}_{\mathrm{ic}}$, we approximate the distribution of $I_{r_{\rmm}}$ with the Gamma distribution, using second-order moment matching \cite{Heath2011}.  The Gamma approximation yields more tractable expressions for $\bbE\{\log(I_{r_\rmm}) \}$  than the challenging to compute density and Laplace characterizations of $I_{r_\rmm}$. 

\begin{remark} (\emph{Gamma Distribution Second Order Moment Matching})\label{rem:gammaapprox}
For the second-order moment matching of the Gamma random variable,  consider a random variable $x$ with finite first $\bbE\{ x\}$ and second order $\bbE\{x^2\}$ moments, and variance $\mathrm{var}(x) = \bbE\{x^2\} -  (\bbE\{ x\})^2$. The Gamma distribution $\Gamma[k,\theta]$ with the same first and second order moments as $x$ has parameters $k = \frac{ (\bbE\{ x\})^2}{\mathrm{var}(x) }$, and $\theta = \frac{\mathrm{var}(x) }{\bbE\{ x\}}$.  
\end{remark}

For the residual intra-cluster  interference, we invoke Jensen's inequality on the first term on the right hand side of (\ref{eqn:deltaTint}) to derive a closed form expression for $\bbE\{I_{\mathrm{res}} \}$

\footnotesize\begin{align}
\nonumber &\bbE\{I_{\mathrm{res}} \}= \bbE\bigg\{\sum_{\ell = 1}^{N}{{(1+r_{0,\ell})^{-\alpha}}|\bg^*_{0,\ell}\widehat{\bff}_\ell|^2}\bigg\}\\
 \nonumber &=  \bbE_{N,r_{0,\ell}}\bigg\{\sum_{\ell = 1}^{N}{{(1+r_{0,\ell})^{-\alpha}}\bbE\bigg\{|\bg^*_{0,\ell}\widehat{\bff}_\ell|^2\bigg\}} \bigg\}\\
\nonumber &\stackrel{(a)}=\bbE_{N,r_{0,\ell}}\bigg\{\sum_{\ell = 1}^{N}{{(1+r_{0,\ell})^{-\alpha}}\frac{\Nt}{\Nt-1} 2^{\rmB_{0,\ell}}\beta\left(2^{\rmB_{0,\ell}}, \frac{\Nt}{\Nt - 1} \right)} \bigg\} \bigg\}
\end{align}
\begin{align}
\nonumber &\stackrel{(b)}=  \bbE_{N,r_{0,\ell}}\bigg\{\sum_{\ell = 1}^{N}{{(1+r_{0,\ell})^{-\alpha}}\Gamma\left(\frac{2\Nt-1}{\Nt - 1}\right) 2^{-\frac{\rmB_{0,\ell}}{\Nt-1}}} \bigg\} \bigg\},
\end{align}\normalsize
where $(a)$ follows from using RVQ for quantization \small{$\bbE\left\{|\bg^*_{0,\ell}\widehat{\bff}_\ell|^2\right\}  = \frac{\Nt}{\Nt-1} 2^{\rmB_{0,\ell}}\beta\left(2^{\rmB_{0,\ell}}, \frac{\Nt}{\Nt - 1} \right)$}\normalsize, \cite{bhagavatula2011}, \cite{Jindal2006}. (b) follows from Stirling's approximation on the Beta function. 
\begin{Result}\label{lem:deltaic}
The mean loss in rate from limited feedback with ICIN, with equal bit allocation, is bounded by

\footnotesize\begin{align}\label{eqn:deltaicebt}
\nonumber& \Delta\widehat{\tau}_{\mathrm{ic}} \leq \log_2(e)\sum_{n=0}^{\infty}{\Gamma\left(\frac{\Nt}{\Nt-1} \right)2^{-\frac{\rmB_{\mathrm{tot}}-n\lfloor\frac{\rmB_{\mathrm{tot}}}{n+1}\rfloor}{{\Nt-1}}}P_N(n)} \\
\nonumber &~-~\log_2(e)\psi\left(k \right) - \log_2(\theta) + \log_2\bigg(\invSNR + k \theta \\
\nonumber &~+~\sum_{n=0}^{\infty}{n\;\Gamma\left(\frac{2\Nt-1}{\Nt-1}\right)2^{-\lfloor\frac{\rmB_{\mathrm{tot}}}{n+1}\rfloor\frac{1}{\Nt-1}}P_N(n)}\bbE\{(1+r_{0,1})^{-\alpha}\} \bigg),
\end{align}\normalsize
where \footnotesize $$\bbE\bigg\{(1+r_{0,1})^{-\alpha}\bigg\} = \displaystyle\int_{0}^{\infty}{\int_{r_0}^{\infty}{(1+r)^{-\alpha} \frac{r}{2\pi\lambda_\rmb}e^{-\frac{r^2}{4\pi\lambda_\rmb}}f_{r_0}(r_0)\rmd r \rmd r_0}},$$\normalsize and $I_{\mathrm{out}} \sim \Gamma[k,\theta]$ via second-order moment matching \cite{Heath2011}, with parameters $k$ and $\theta$ such that $\bbE\{I_{r_\rmm}\} = k\theta$ and $\mathrm{var}( I_{r_\rmm}) = k\theta^2$, with \\$\bbE\{I_{r_\rmm}\} = 2\pi\lambda_\rmb\;\bbE_{r_\rmm, r_0}\left\{\int_{r_\rmm - r_0}^{\infty}{\frac{r}{L(r)}\rmd r}\right\}$ and
$\mbox{var}(I_{r_\rmm}) = 2\pi\lambda_\rmb\; \bbE_{r_\rmm, r_0}\left\{\int_{r_\rmm - r_0}^{\infty}{\frac{r}{(L(r))^2}\rmd r}\right\}$.
\end{Result}
\begin{IEEEproof}
See Appendix \ref{app:app_4}.
\end{IEEEproof}
The mean loss in rate $\Delta\widehat{\tau}_{\mathrm{ic}}$ decreases as the total number of bits $\rmB_{\mathrm{tot}}$ increases. $\Delta\widehat{\tau}_{\mathrm{ic}}$ increases, however, with the average cluster size  $\lambda_{\rmb}/\lambda_{\rmc}$, given a fixed feedback budget $\rmB_{\mathrm{tot}}$. A higher  $\lambda_\rmb/\lambda_\rmc$ implies more base stations per cluster and hence more interfering channels to quantize. As the number of antennas at the base stations increases with the number of base stations per cluster for ICIN to be feasible, the number of bits allocated per antenna for each quantized channel also decreases. 

\subsection{Adaptive Bit Allocation}
As the equal bit allocation limited feedback strategy results in a considerable decrease in the ICIN performance, as will be illustrated in Section \ref{sec:simRes}, in this section we optimize the bit allocation to minimize the mean loss in achievable rate of ICIN. We adapt the number of  bits allocated to the desired and interfering channels as a function of their signal strength at the typical user. Since each spatial realization results in a different number of intra-cluster interferers $N$, the optimization is done per spatial realization, as a function of $N$. 

We denote the total number of bits allocated to quantizing the interfering channels by $\rmB_i = \rmB_{\mathrm{tot}} - \rmB_0$. Given $\rmB_i$, we first derive $\rmB_{0,\ell}, \ell = 1,\cdots,N$ such that the contribution of the interfering channels towards the mean loss in rate is minimized. In other words, we aim at finding  $\rmB_{0,\ell}$ such that $I_{\mathrm{res}} $ is minimized. 
The optimization problem is expressed as

\footnotesize\begin{eqnarray}\label{eqn:opt}
&\displaystyle\min_{\rmB_{0,1},\cdots,\rmB_{0,N}}~\sum_{\ell = 1}^{N}{{(1+r_{0,\ell})^{-\alpha}}\Gamma\left(\frac{2\Nt-1}{\Nt - 1}\right) 2^{-\frac{\rmB_{0,\ell}}{\Nt-1}}}\\ 
\nonumber &\mbox{s.t. }\sum_{\ell=1}^{N}{\rmB_{0,\ell}} = \rmB_i, \mbox{ and } \rmB_{0,\ell} > 0.
\end{eqnarray}\normalsize
The bit allocations $\rmB_{0,\ell}, \ell \in 1\cdots,N$ are integer-valued. We solve a relaxation of the optimization problem assuming $\rmB_{0,\ell}$ are real valued. The solution is given by the arithmetic-geometric mean inequality. It is derived in \cite{bhagavatula2011} and restated here for completeness.
\begin{Lemm}{(\cite[Theorem 4]{bhagavatula2011})}
The optimum number of bits assigned to the $\ell$-th interferer, $\rmB^*_{0,\ell}$, that minimizes  (\ref{eqn:opt}) is given by

\small\begin{eqnarray}\label{eqn:solBi}
\rmB^*_{0,\ell} = \frac{\rmB_i}{|\cK|} + (\Nt-1)\log_2\left(\frac{(1 + r_{0,\ell})^{-\alpha}}{\prod_{\ell\in\cK}{(1 + r_{0,\ell})^{-\alpha/|\cK|}}} \right)
\end{eqnarray}\normalsize
for $\ell \in \cK$ and $\rmB_{0,\ell} = 0$ for $\ell \notin \cK$, where $\cK$ is the largest set of interferers that satisfies
\begin{eqnarray}
\log_2\left(\frac{\prod_{\ell\in\cK}{(1 + r_{0,\ell})^{-\alpha/|\cK|}}}{(1 + r_{0,\ell})^{-\alpha}} \right) < \frac{\rmB_{i}}{|\cK|(\Nt-1)}.
\end{eqnarray}
$\cK$ denotes the set of effective interferers, such that $|\cK| \leq N$.
\end{Lemm}
Using the  expression for $\rmB_{0,\ell}$ from (\ref{eqn:solBi}), the bound on the mean loss in rate per spatial realization, i.e. given $N$ and $r_{0,\ell}$ is expressed as a function of a single variable $\rmB_0 = \rmB_{\mathrm{tot}} - \rmB_i$, 

\footnotesize\begin{align}
\nonumber &\Delta{\widehat{\widehat{\tau}}_{\mathrm{ic}}} (\rmB_0)\approx \log_2(e)\Gamma\left(\frac{\Nt}{\Nt-1}\right)2^{-\frac{\rmB_{0}}{\Nt-1}}\\
\nonumber &-  \bbE\bigg\{\log_2\left({I_{\mathrm{out}}+ \invSNR}\right) \bigg\}\\
\nonumber &+ \log_2\left(\Gamma\left(\frac{2\Nt-1}{\Nt-1}\right)|\cK|2^{-\frac{\rmB_{\mathrm{tot}} - \rmB_0}{|\cK|(\Nt-1)}}\prod_{\ell\in\cK}{(1 + r_{0,\ell})^{-\alpha/|\cK|}}\right.\\ &\left.  \qquad\qquad\qquad + \bbE\{I_{\mathrm{out}}\} + \invSNR\right).
\end{align}\normalsize
To minimize the mean loss in rate, we minimize $\Delta{\widehat{\widehat{\tau}}_{\mathrm{ic}}} (\rmB_0)$ as a function of $\rmB_0$ such that $\rmB_0 \leq \rmB_{\mathrm{tot}}$. We distinguish between two cases of interest, depending on the ratio of residual interference  $I_{\mathrm{res}}$ to inter-cluster interference $I_{\mathrm{out}}$.

\textbf{Dominant Inter-Cluster Interference}: When \small$\bbE\{I_{\mathrm{res}}\} < \bbE\{I_{\mathrm{out}}\}+1/\mathsf{SNR}$\normalsize, we use the low-SNR approximation  $\log(1+x) \approx x$. The bound on the mean loss in rate is written as

\footnotesize\begin{align}\label{eqn:meanratesmall}
\nonumber &\Delta{\widehat{\widehat{\tau}}_{\mathrm{ic}}} (\rmB_0)\approx \log_2(e)\Gamma\left(\frac{\Nt}{\Nt-1}\right)2^{-\frac{\rmB_{0}}{\Nt-1}}\\
\nonumber &-  \bbE\bigg\{\log_2\left({I_{\mathrm{out}}+ \invSNR}\right) \bigg\} +\log_2(\bbE\{I_{\mathrm{out}}\}+ \invSNR)\\
 &+ \log_2(e) \frac{\Gamma\left(\frac{2\Nt-1}{\Nt-1}\right)}{\bbE\{I_{\mathrm{out}}\} + \invSNR} |\cK|2^{-\frac{\rmB_{\mathrm{tot}} - \rmB_0}{|\cK|(\Nt-1)}}\prod_{\ell\in\cK}{(1 + r_{0,\ell})^{-\alpha/|\cK|}}.
\end{align}\normalsize
The optimization is carried out for the terms in $\rmB_0$. 
\begin{Result} \label{res:DominantInterCluster}
 The optimal value of $\rmB_0$ for \small$\bbE\{I_{\mathrm{res}}\} < \bbE\{I_{\mathrm{out}}\}+1/\mathsf{SNR}$\normalsize, given $N$ and $r_{0,\ell},\ell = 1\cdots N$, is

\footnotesize\begin{eqnarray}\label{eqn:B_0IresIout}
\nonumber\rmB_0 &=& \frac{\rmB_{\mathrm{tot}}}{|\cK|+1} - \log_2\left(\frac{\Nt |\cK|}{\Nt-1}\prod_{\ell=1}^{|\cK|}{(1 + r_{0,\ell})^{-\frac{\alpha}{|\cK|}}}\right) \frac{(\Nt-1)|\cK|}{|\cK|+1} \\
&+&\log_2(\bbE\{I_{\mathrm{out}}\} + \invSNR)\frac{(\Nt-1)|\cK|}{|\cK|+1}.
\end{eqnarray}\normalsize
\end{Result}
\begin{IEEEproof}
The proof is provided in Appendix \ref{app:app_5}.
\end{IEEEproof}
\textbf{Dominant Residual Intra-Cluster Interference} When \small$\bbE\{I_{\mathrm{res} }\}> \bbE\{I_{\mathrm{out}}\} + 1/\mathsf{SNR}$\normalsize, we use the high SNR approximation $\log(1+x) \approx \log(x)$ and we write $\Delta{\widehat{\widehat{\tau}}_{\mathrm{ic}}} (\rmB_0)$ as

\footnotesize\begin{align}\label{eqn:highSNR}
 \nonumber\Delta{\widehat{\widehat{\tau}}_{\mathrm{ic}}} (\rmB_0) \approx \log_2(e)\Gamma\left(\frac{\Nt}{\Nt-1}\right)2^{-\frac{\rmB_{0}}{\Nt-1}} -\bbE\bigg\{\log_2\left({I_{\mathrm{out}}+ \invSNR}\right) \bigg\} \\
+\log_2\left(\Gamma\left(\frac{2\Nt-1}{\Nt-1}\right)|\cK|\prod_{\ell\in\cK}{(1 + r_{0,\ell})^{-\alpha/|\cK|}}\right) + \frac{\rmB_0-\rmB_{\mathrm{tot}}}{|\cK|(\Nt-1)}.
\end{align}\normalsize
Since the terms which are a function of $\rmB_0$ in (\ref{eqn:highSNR}) are convex with respect to $\rmB_0$, the optimal value of $\rmB_0$ is found by taking the derivative of these terms with respect to $\rmB_0$.
\begin{Result}
The optimal value of $\rmB_0$ for \small$\bbE\{I_{\mathrm{res} }\}> \bbE\{I_{\mathrm{out}}\}+1/\mathsf{SNR}$\normalsize, given $N$, is

\footnotesize\begin{eqnarray}\label{eqn:B0highSNR}
\rmB_0 = (\Nt-1)\log_2\left(|\cK|\log_2(e)\Gamma\left(\frac{\Nt}{\Nt-1}\right)\right).
\end{eqnarray}\normalsize
\end{Result}
The optimization problems for both dominant inter-cluster and dominant intra-cluster interference are solved assuming non-integer bit values. Since the objective functions in (\ref{eqn:meanratesmall}) and (\ref{eqn:highSNR}) are convex in $\rmB_0$, we get the integer value of $\rmB_0$ by taking the floor or the ceiling of the value in $(\ref{eqn:B_0IresIout})$ and (\ref{eqn:B0highSNR}), respectively. 

An upper bound on mean loss in rate with adaptive bit allocations, averaged over all spatial realizations, is subsequently given by replacing $\rmB_0$ by its optimal value  in (\ref{eqn:meanratesmall}), distinguishing between the two dominant interference cases.

\section{Performance Evaluation with Perfect CSI with Fixed $\Nt$}\label{sec:thresh}
In the analysis so far, we assumed that $\Nt$ is a function of the number of interferers in the cluster and hence is a random variable. This assumption makes the analysis with perfect and imperfect CSI more tractable, and allows system designers to gauge how many antennas they should provide at the base stations for interference coordination to be most beneficial. 

In this section, since the number of antennas may be predetermined, we relax this assumption and we assume $\Nt$ is fixed at each base station. As $N$ is a random variable that changes per spatial realization and can become greater than $\Nt$, interference coordination may no longer be feasible per cluster. To overcome this issue, we propose a thresholding policy wherein if $\Nt \leq N$, each base station beamforms to its own user and no interference coordination is performed. When $\Nt > N$, ICIN is applied as in Section \ref{sec:perfEval}. 

Let $A$ be the event that the number of interferers in the cluster is less than $\Nt$, then for a fixed $\Nt$ the probability of coverage is expressed as

\footnotesize\begin{eqnarray}
\nonumber p_\rmc(\lambda_\rmb,\lambda_\rmc,\alpha,\Nt) &=&{\bbP\left[\mathsf{SIR}_0\geq T | A\right]}\bbP\left[A\right] + {\bbP\left[\mathsf{SIR}_0\geq T | A^c\right]}\bbP\left[A^c\right]\\
&=&\underbrace{\bbP\left[\mathsf{SIR}_0\geq T \cap A\right]}_{p_{\rmc,\mathrm{ic},\Nt}} +\underbrace{\bbP\left[\mathsf{SIR}_0\geq T \cap A^c\right]}_{p_{\mathrm{c},\mathrm{nic},\Nt}}.
\end{eqnarray}\normalsize
The probability of coverage with per cluster coordination and $\Nt> N$ follows from the probability of coverage expression derived in Section \ref{sec:perfEval} with one difference being that the averaging is now done over $N$, given that $\Nt$ is fixed and the Laplace transform of the desired signal depends on $\Nt - N$, now a random variable. Using the PMF of $N$  in Section \ref{sec:impLim}, the lower bound on the probability of coverage $p_{\rmc,\mathrm{ic},\Nt}$ is 

\footnotesize\begin{align}
\nonumber &p_{\rmc,\mathrm{ic},\Nt}(\lambda_\rmb,\lambda_\rmc, \alpha, T) \geq \sum_{n=0}^{\Nt-1}\left\{p_N(n)\int_{0}^{\infty}{f_r(r_0)\displaystyle\int_{r_0}^{\infty}f_{r_\rmm}(r_\rmm)}\times\right.\\
\nonumber &\left.\int_{-\infty}^{\infty}{e^{-2\pi j\frac{L(r_0)T}{\mathsf{SNR}} s}\cL_{I_{r_\rmm }}\left(2j\pi L(r_0)T s\right)\frac{\cL_h(-2j\pi s)-1}{2i\pi s}\rmd s}\rmd r_0 \rmd r_\rmm \right\},
\end{align}\normalsize
where the desired signal $|\bh^*_0\bff_0|^2 \sim \Gamma[\Nt-n,1]$, and $p_N\left[n\right]$, $f_c(c)$ and $f_r(r_0)$ are given in (\ref{eqn:p_N}) and Result \ref{lem:pic}, respectively.

With single-cell beamforming and no interference coordination, the interference at the typical user $\rmu_0$ is divided into two independent components, the inter-cluster interference $I_{\mathrm{out}}$ bounded by $I_{r_\rmm}$, and the intra-cluster interference denoted by $I_{\mathrm{in}}$. The SINR, with perfect CSI beamforming, is given by
\begin{figure*}
\footnotesize\begin{eqnarray}
\mathsf{SINR}_{\mathrm{nic},\Nt} = \frac{\gamma_0\|\bh_0\|^2}{\sum_{\rmb_\ell \in \Pi_{\rmb}/V^{(\rmc)}(\rmb_0)}{\gamma_{0,\ell}|\bg^*_{0,\ell}\bff_\ell|^2} + \sum_{\rmb_\ell \in V^{(\rmc)}(\rmb_0)}{\gamma_{0,\ell}|\bg^*_{0,\ell}\bff_\ell|^2} + \sigma^2} =  \frac{(1+ r_0)^{-\alpha}\|\bh_0\|^2}{I_{\mathrm{out}}+ I_{\mathrm{in}} + \frac{1}{\mathsf{SNR}}},
\end{eqnarray}\normalsize
\end{figure*}
where the desired signal power $\|\bh_0\|^2 \sim \Gamma[1,\Nt]$, and $I_{\mathrm{out}}$ and $I_{\mathrm{in}}$ are independent by the PPP property, since the interfering base stations are located  in two disjoint areas. 

The probability of coverage with beamforming, with $\Nt <N$, is expressed as

\footnotesize\begin{align}
\nonumber &p_{\rmc,\mathrm{nic},\Nt}(\lambda_\rmb,\lambda_\rmc, \alpha, T)= \\
&\bbE_{N,{r_0}}\bigg\{\bbP\left[\|\bh\|^2 > L(r_0)T(I_{\mathrm{out}}+ I_{\mathrm{in}} + \invSNR)\right]\bigg\}.
\end{align}\normalsize
Conditioned on the value of $N$, the number of interferers inside the cluster is fixed, and the point process corresponding to the base stations interferers inside the cluster forms a binomial point process \cite{Stoyan1995}. The intra-cluster interference term $I_{\mathrm{in}}$ is thus a result of a binomial point process in the area formed by the cluster.
For the contribution of $I_{\mathrm{in}}$, we compute an upper bound assuming the interference is the aggregation of the signal powers of the interfering base stations in the annular region with inner radius $r_0$ and outer radius $r_\rmM$ corresponding to the circumscribed circle to the cluster. 

The probability of coverage with no interference nulling is 

\footnotesize\begin{align}
\nonumber &p_{\rmc,\mathrm{nic},\Nt}\geq \left\{\displaystyle\sum_{n=\Nt}^{\infty}{\bbP\left[N=n\right]\displaystyle\int_{0}^{\infty}{f_{r_0}(r_0)\displaystyle\int_{r_0}^{\infty}f_{r_\rmm}(r_\rmm)}\displaystyle\int_{r_0}^{\infty}f_{r_\rmM}(r_\rmM)}\right.\\
\nonumber&\qquad\qquad\times\left.{{{\int_{-\infty}^{\infty}{e^{-2\pi j s/\mathsf{SNR}}\cL_{I_{r_\rmm}}(2j\pi s)\cL_{I_{r_\rmM}}(2j\pi s)}}}}\right.\\
\nonumber&\qquad\qquad\qquad\qquad\left.{{{{\frac{\cL_h(-2j\pi (L(r_0)T)^{-1} s)-1}{2j\pi s}\rmd s}\rmd r_\rmM } \rmd r_\rmm \rmd r_0} }\right\},
\end{align}\normalsize
where $r_\rmM$ is such that its CCDF is bounded by, \cite{Calka2002}

\footnotesize\begin{eqnarray}
\nonumber\bbP[r_M >r] \geq 2\pi r^2 e^{-\pi r^2} (1 + \frac{1}{2\pi r^2}e^{-\pi r^2})
\end{eqnarray}\normalsize
and the Laplace transform of $I_{r_\rmM}$ is given by

\footnotesize\begin{eqnarray}
\nonumber & L_{I_{r_\rmM}}(s) = \bigg\{1 - \frac{2\pi ((r_0+1)(r_M + 1))^{-\alpha}s}{\pi r_M^2 - \pi r_0^2}\times 
\bigg\{ \\\nonumber & \frac{1}{\alpha-2}\left((r_0+1)^{2}(r_\rmM + 1)^\alpha {}_2F_1(1,1-\frac{2}{\alpha}, 2- \frac{2}{\alpha}, -(r_0+1)^{-\alpha}s)\right.\\
&\nonumber\left. - (r_\rmM + 1)^{2}(r_0+1)^\alpha {}_2F_1(1,1-\frac{2}{\alpha},2-\frac{2}{\alpha},-(r_\rmM+1)^{-\alpha}s)\right)\\
\nonumber &~+~   \frac{1}{\alpha-1}\left(-(r_0+1)(r_\rmM + 1)^\alpha {}_2F_1(1,1-\frac{1}{\alpha}, 2- \frac{1}{\alpha}, -(r_0+1)^{-\alpha}s)\right.\\
&\nonumber\left. (r_\rmM + 1)(r_0+1)^\alpha {}_2F_1(1,1-\frac{1}{\alpha},2-\frac{1}{\alpha},-(r_\rmM+1)^{-\alpha}s)\right)\bigg\}\bigg\}^N.
\end{eqnarray}\normalsize
Combining the expressions for $p_{\rmc,\mathrm{ic},\Nt}$ and $p_{\rmc,\mathrm{nic},\Nt}$, we obtain the main result on the probability of coverage $p_{\rmc,\Nt}(\lambda_\rmb,\lambda_\rmc, \alpha, T)$  with interference coordination, for fixed $\Nt$ with thresholding. 

The average throughput for fixed $\Nt$ follows from the probability of coverage analysis similarly to Section \ref{sec:perfEval}.

\section{Simulation Results and Discussion}\label{sec:simRes}
We consider a  surface comprising on average $100$ clusters. For random clustering, the density of the cluster base stations $\lambda_\rmc$ is varied such that ${\lambda_\rmc}/{\lambda_\rmb} \geq 1$ to show the benefits of increasing cluster sizes.  The path-loss exponent is set to $\alpha = 4$.

Throughout the numerical evaluation, we compare the performance of per cluster ICIN with that of unconditional beamforming, where each base station beamforms to its own users, irrespective of the number of antennas $\Nt$ at the base stations. With unconditional beamforming, the typical mobile user is subject to interference from all the other base stations in the network, and the interference power is a shot noise process from all the interferers. 
\begin{figure}[t]
  \begin{center}
  \vspace{-10pt}
    \includegraphics[scale = 0.37]{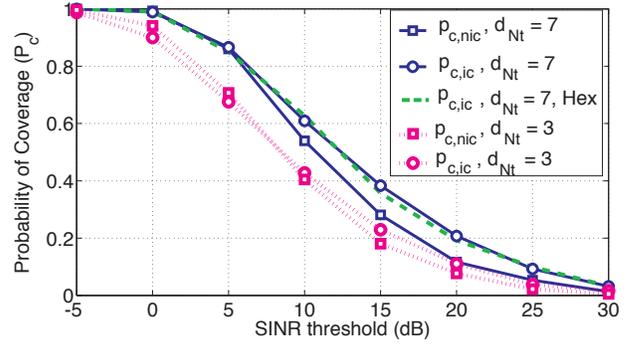}
    \caption{The probability of coverage for increasing SINR threshold. $p_{\rmc,\mathrm{ic}}$ denotes the coverage with per cluster ICIN. $p_{\rmc,\mathrm{nic}}$ denotes the probability of coverage with beamforming and no nulling. The average number of base stations in the cell is $3$ and the pathloss exponent is 4.}    \label{fig:comp_outage}
  \end{center}
\end{figure}

Figure \ref{fig:comp_outage} plots the probability of coverage versus the SINR threshold $T$ for increasing $\rmd_\Nt = \Nt - N$ at the transmitters, and for fixed average cluster size $\lambda_\rmb/\lambda_\rmc = 3$.  The probability of coverage increases with increasing $\rmd_\Nt$,  $\rmd_\Nt = 3, \rmd_\Nt = 7$. For low SINR, and for moderate $\mathsf{SNR}$ single-cell beamforming outperforms ICIN for $\rmd_\Nt = 3$. ICIN outperforms beamforming as the SINR threshold increases. This can be justified by the tradeoff between the decrease in the signal power for ICIN with the decrease in the interference power at low SINR. At higher SINR values, the system is more interference limited, and decreasing the interference when using ICIN outperforms boosting the signal power when using beamforming. Moreover,  for ICIN, as the clustering is done at random, when the base station is at the edge of the cluster, the interference from adjacent clusters remains dominant, leading to a smaller decrease in the interference power and hence lower ICIN coverage gains. For $\rmd_\Nt = 3$, ICIN outperforms no nulling in coverage starting at $T = 7$ dB. This threshold decreases with increasing $\rmd_\Nt$. For $\rmd_\Nt = 7$, ICIN outperforms no nulling for all the $T$ values of interest. Figure \ref{fig:comp_outage} also plots the coverage probability performance for fixed lattice grid clustering proposed in \cite{Huang2012j}, for $\rmd_\Nt = 7$. We can conclude from the comparison that the two models are equivalent in terms of coverage performance. 
%

\begin{figure}[t]
  \begin{center}
  \vspace{-10pt}
    \includegraphics[scale = 0.37]{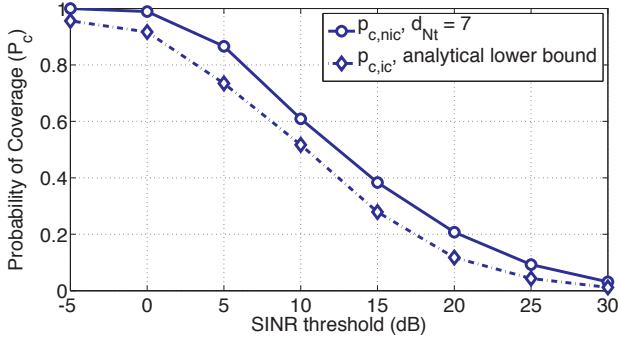}
    \caption{The probability of coverage for increasing SINR threshold. $p_{\rmc,\mathrm{ic}}$  obtained by simulations is compared to the analytical lower bound derived in Result 1. The average number of base stations in the cell is $3$ and the pathloss exponent is 4. }
    \label{fig:comp_outage_bound}
  \end{center}
\end{figure}
Figure \ref{fig:comp_outage_bound} compares the probability of coverage obtained from Monte Carlo simulations with the bound derived in Result 1 for $\rmd_\Nt = 7$ and average cluster size equal to $3$. Figure \ref{fig:comp_outage_bound}  shows that the lower bound on the probability of coverage exhibits the same behavior as the Monte Carlo simulation. It is sufficiently accurate, to within 0.1 in probability, and can provide insights on the performance of the clustered coordination system. 

\begin{figure}[t]
  \begin{center}
  \vspace{-10pt}
    \includegraphics[scale = 0.33]{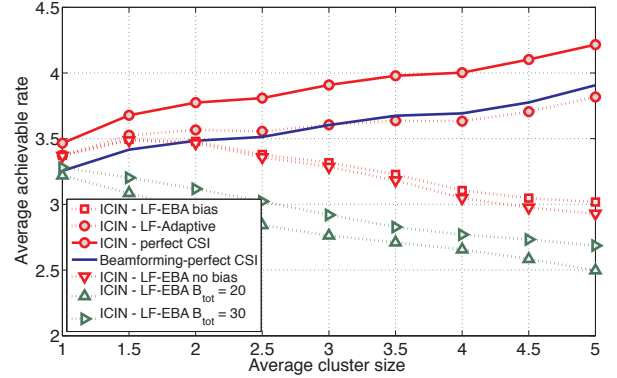}
    \caption{The average  rate as a function of  average cluster size, for $\Nt = N + 4$. $\tau$ (with no ICIN) also shown for comparison. The feedback budget per mobile user is fixed $\rmB_{\mathrm{tot}} = 50$ for all the curves except the LF-EBA curves with $\rmB_{\mathrm{tot}} = \{20, 30\}$ indicated in the legend. The performance of beamforming (no ICIN) with limited feedback is almost equivalent to the performance with perfect CSI. LF-EBA-Bias denotes the equal bit allocation scheme with extra bits resulting from rounding to an integer bit value given to the desired user. LF-EBA-no Bias denotes the equal bit allocation scheme with the extra bits not used.}
    \label{fig:comp_avgclustersize}
  \end{center}
\end{figure}

Figure \ref{fig:comp_avgclustersize} plots the average  rate obtained with ICIN as a function of average cluster size, for $\rmd_\Nt = 4$. It compares the average  rate with equal bit allocation limited feedback (LF-EBA) with and without bias for the desired channel, and the average rate with adaptive limited feedback. The average  rate with no interference nulling is also shown for comparison.
For a fixed feedback budget at each receiver of $\rmB_{\mathrm{tot}} = 50$, Figure \ref{fig:comp_avgclustersize} shows that as the average cluster size increases, the average  rate with ICIN increases. This is due to the decrease in the interference terms as more and more base stations are added to the cluster. The same holds for the beamforming strategy, as the increase in the number of antennas at each base station as a function of the cluster size increases the signal term, leading to an increase in SINR. For adaptive limited feedback, the average  rate increases with the average cluster size, similarly to the increase of ICIN with perfect CSI. ICIN with adaptive limited feedback outperforms no ICIN for moderate average cluster sizes up to $\lambda_\rmb/\lambda_\rmc = 4$. It performs almost similarly to no ICIN for larger cluster sizes.
For limited feedback beamforming with equal bit partitioning however, the average rate increases, reaches a maximum at $\lambda\rmb/\lambda_\rmc = 2$, then decreases. This is justified by the double increase in quantization error due to the decrease in the number of bits per channel and decrease in the number of bits per antenna. Limited feedback with equal bit partitioning and biasing towards the desired channel, through allocating the extra bits available after rounding the bit allocations to integer values, as discussed in Section V, performs better than limited feedback with equal bit partitioning and no desired channel biasing. The extra bits available at the mobile station from the remainder of the division of $\rmB_{\mathrm{tot}}$ by $N+1$ are more judiciously used to quantize the desired signal in the equal bit allocation with bias. 
To further illustrate the tradeoff of the average rate with equal bit partitioning with the average cluster size, we plot the average rate performance of LF-EBA with biasing for $\rmB_{\mathrm{tot}} = \{20, 30\}$ in the same figure. For low feedback budget, the maximum is achieved at an average cluster size of 1. 
\begin{figure}[t]
  \begin{center}
  \vspace{-10pt}
    \includegraphics[scale = 0.33]{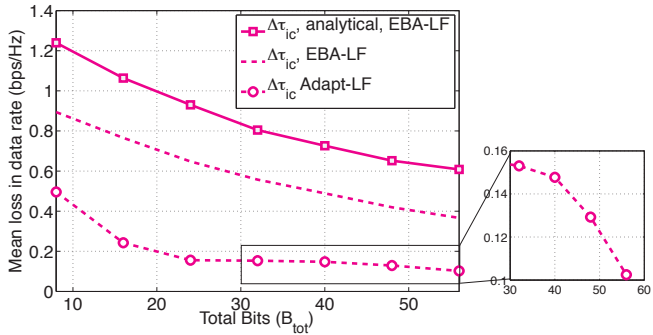}
    \caption{The mean loss in rate versus the feedback budget per mobile $\rmB_{\mathrm{tot}}$ is shown. $\Nt - N=5$. and average cluster size is fixed at $\lambda_\rmb/ \lambda_\rmc = 3$.}
    \label{fig:comp_bits}
  \end{center}
\end{figure}

Figure \ref{fig:comp_bits} plots the mean loss in rate versus the total number of bits $\rmB_{\mathrm{tot}}$ available at the receiver for $\lambda_\rmb/\lambda_\rmc =3$, and $\rmd_\Nt = 5$. The figure also shows the analytical upper bound derived in Result 3 for comparison. The bound is sufficiently tight for all values of $\rmB_{\mathrm{tot}}$. The gap is due to the bounds on the quantization errors in the RVQ analysis as well as the Jensen's inequality and the Gamma second order moment matching. Limited feedback with adaptive bit allocation is shown to significantly decrease the mean loss in rate, as compared to equal bit allocation. To show the decrease of the mean loss in rate for adaptive limited feedback for large values of $\rmB_{\mathrm{tot}}$, we zoom in on $\rmB_{\mathrm{tot}} > 30$. The mean loss in rate with adaptive limited feedback decreases as the number of bits increase, albeit at a lower rate than that with equal bit partitioning. 

\begin{figure}[t]
  \begin{center}
  \vspace{-10pt}
    \includegraphics[scale = 0.37]{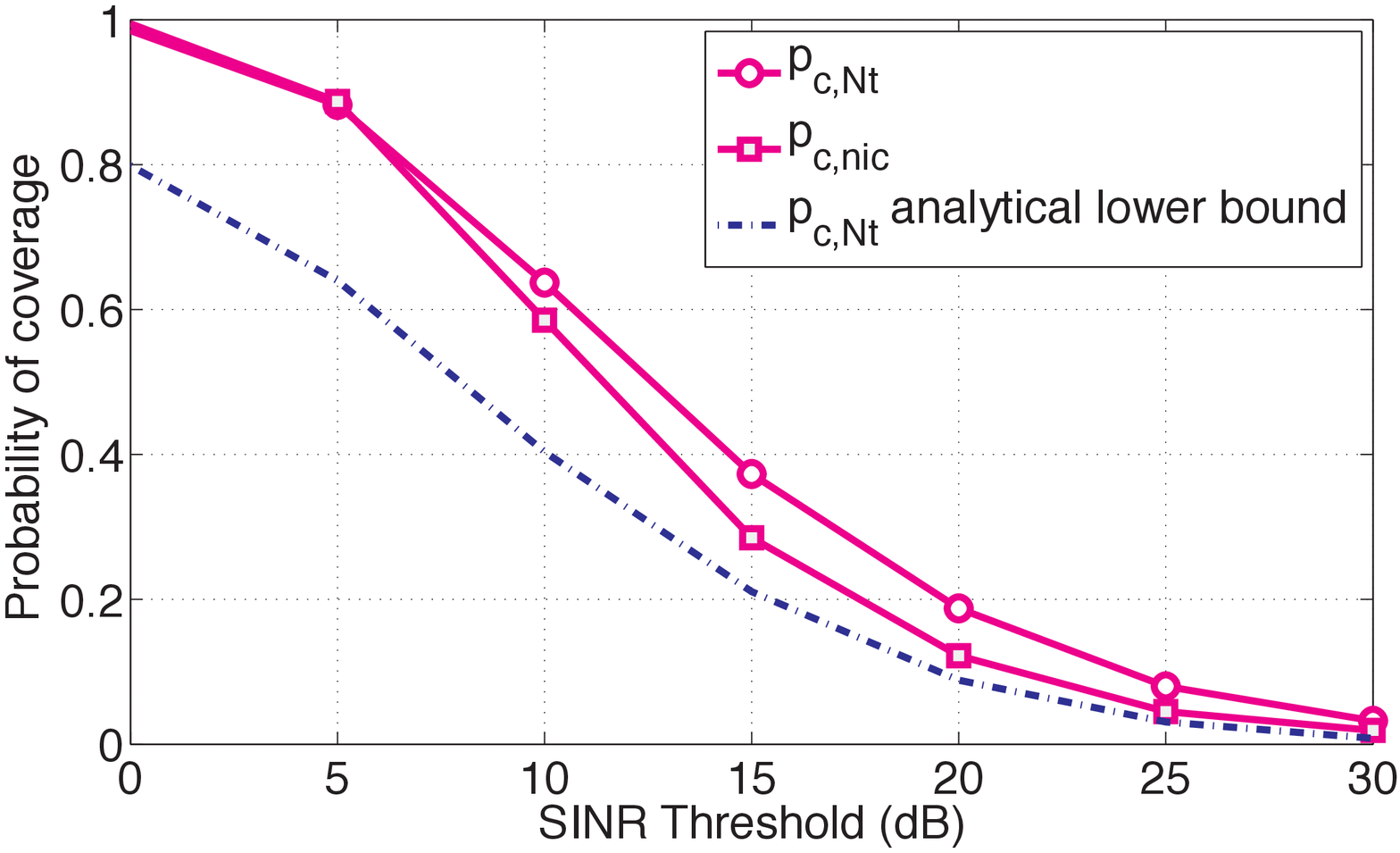}
    \caption{The probability of coverage for increasing SINR threshold. $p_{\rmc,\mathrm{ic},\Nt}$  obtained by simulations is compared to the analytical lower bound derived in Result 1. The average number of base stations in the cell is $3$ and the pathloss exponent is 4. }
    \label{fig:comp_bound_tresh}
  \end{center}
\end{figure}
Figure \ref{fig:comp_bound_tresh} compares the probability of coverage obtained from Monte Carlo Simulations with the bound derived in Section VI,  for $\Nt = 10$ and average cluster size equal to $3$. The Figure shows that the lower bound on the probability of coverage exhibits the same behavior as the Monte Carlo simulation. The bound, however is loose for small SINR threshold values. This is due to using the circumscribed and inscribed circle to bound the interference power for coordination and no coordination. 

\begin{figure}[t]
  \begin{center}
  \vspace{-10pt}
    \includegraphics[scale = 0.42]{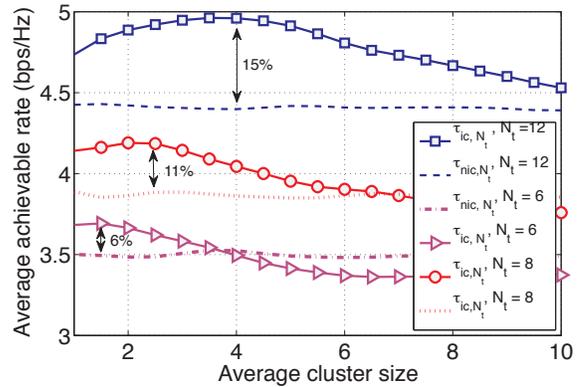}
    \caption{ The average  rate $\tau$ as a function of the average cluster size, for increasing $\Nt = \{6, 8,12\}$. $\tau(\mbox{no ICIN})$  is also shown for comparison. $\tau_{\mathrm{ic},\Nt}$ increases with $\lambda_\rmb/\lambda_\rmc$, reaches a maximum and then decreases. The gains from ICIN depend on the number of antennas $\Nt$ at the base stations.} 
    \label{fig:comp_threshold}
  \end{center}
\end{figure}
Figure \ref{fig:comp_threshold} plots the average  rate obtained with ICIN using the thresholding policy in Section VI, as a function of average cluster size, for increasing number of antennas $\Nt = \{6, 8, 12\}$. The achievable rate with unconditional beamforming $\tau_{\mathrm{nic},\Nt}$ is also shown for comparison. $\tau_{\mathrm{nic},\Nt}$ does not depend on the average cluster size; it is thus almost constant for all values of $\lambda_\rmb/\lambda_\rmc$. 
$\tau_{\mathrm{ic}, \Nt}$ increases with increasing $\lambda_\rmb/\lambda_\rmc$, reaches a maximum, then decreases towards $\tau_{\mathrm{nic}, \Nt}$.  This is due to the thresholding policy where the feasibility of ICIN depends on the number of interferers per cluster. When $\lambda_\rmb/\lambda_\rmc$ is large compared to the number of antennas, ICIN feasibility decreases, and hence single-user beamforming becomes more prevalent. The gains from coordination depend on the number of antennas at the base stations and the average cluster size. For $\Nt=12$ for example, as the average cluster size increases, the average  rate increases, reaches a maximum at around $\lambda_\rmb/\lambda_\rmc = 4$ and then decreases, until it reaches $\tau_{\mathrm{nic},\Nt}$. As the number of antennas increases, ICIN feasibility increases, and the average cluster size where ICIN with thresholding attains maximum gains increases. 
The maximum of $\tau_{\mathrm{ic,\Nt}}$ depends on the average cluster size, the number of antennas $\Nt$ and the variance of the cluster size. The variance of the cluster size is a function of the densities $\lambda_\rmb$ and $\lambda_\rmc$. As $\Nt$ increases, $\tau_{\mathrm{ic,\Nt}}$ increases, and the relative gain from interference coordination increases. It reaches $15\%$ for $\Nt = 12$ and $8\%$ for $\Nt = 6$.

\section{Conclusion}\label{sec:conc}
In this paper, we analyzed intra-cluster interference coordination for randomly deployed base stations. To cluster the randomly deployed base stations, we proposed a random clustering strategy that overlays an independent Poisson point process on the base stations point process. We assumed that the base stations connected to the closest cluster center form a cluster. We showed that the average cluster size can be optimized with respect to the number of antennas at the base stations to provide the maximum gains from ICIN. We further analyzed the performance of per cluster ICIN with limited feedback CSI. 
We  showed that limited feedback CSI hinders the gains of coordination, and results in significant loss in rate, when equal bit allocation is used. Adaptive bit allocation, optimized as a function of the signal strengths, recover the gains of clustered coordination.  One takeaway from this paper is that analysis of cooperative multi-cell systems, for randomly deployed base stations, is possible using hierarchical association.  Ongoing work includes using the same setup to analyze the performance of clustered coordination with CSI and user data sharing. It also includes deriving an analytical framework for dynamic user-driven clustering, using the same framework.

\appendices

\section{Proof of Lemma \ref{lem:pic}}\label{app:app_1}
Conditioned on $\rmb_0$ being at a distance $r_0$ from $\rmu_0$, the interference outside a ball of radius   $r_\rmm - r_0 - r_1$ away from $\rmu_0$,
the probability of coverage is given by

\footnotesize\begin{align}
\nonumber &p_{\rmc,\mathrm{ic}} = \\
\nonumber &\bbE_{r_\rmm, r_0, r_1}\left\{\bbP\left[|\bh^*_0\bff_0|^2\geq \frac{T}{(1+ r_0)^{-\alpha}}\left(I_{\rmB(\rmu_0)} + \frac{1}{\mathsf{SNR}}\right) \bigg| r_0, r_{\rmm}, r_1\right]  \right\}\\
\nonumber &\stackrel{(a)}= \int_{0}^{\infty}f_{r_1}(r_1)\int_{0}^{\infty}{f_{r_0}(r_0)\int_{r_0}^{\infty}{f_{r_\rmm}(r_\rmm)}}{{\int_{-\infty}^{\infty}{e^{-\frac{2j\pi s L(r_0) T}{\mathsf{SNR}}}}}}\\
\nonumber &{{{\cL_{I_{r_\rmm}}(2j\pi \ell(r_0) T s)\frac{ \left(\frac{1}{-2j\pi s + 1}\right)^{\Nt-N}-1}{2j\pi s}\rmd s}\rmd r_\rmm}\rmd r_0 \rmd r_1}.
\end{align}\normalsize
$(a)$ is derived using the expression in \cite{baccelli2009} for the probability of coverage using general fading distributions.  This expression is applicable here since the desired signal power  $|\bh^*_0\bff_0|^2$ has a finite first moment and admits a square integrable density. The interference $I_{r_\rmm}$ is square integrable using the path-loss model $(1+r)^{\alpha}$ reminiscent of the pathloss model $(1+Ar)^{\alpha}$ in \cite{Baccelli2009a}. The distribution of $r_1$ is given by $f_{r}(r_1) = 2\pi\lambda_\rmc r_1 e^{-\pi r_1^2\lambda_\rmc}$.  

The aggregate interference considered outside a ball of radius $r_\rmm - r_0 - r_1$ results in an upper bound on the aggregate interference due to the fact that less area is excluded from the calculations. The inter-cluster interference can also be bounded by the interference outside a ball of radius $r_\rmm$ centered at $\rmu_0$, corresponding to a slightly larger area to be excluded from the calculations than $r_\rmm - r_0 - r_1$. This averaged over the spatial realizations, results in a good lower bound on the probability of coverage as illustrated in Section VII.  The lower bound is finally given by the expression in Lemma \ref{lem:pic}. 

\section{Proof of Remark \ref{rem:pN}} \label{app:app_3}
The PMF of $N$ is computed as the Poisson distributed number of base stations inside a cluster of size $t$, such that this cluster contains one base station chosen at random from the process $\Pi_b$. 

To derive the distribution of the size $t$ of the cluster conditioned on the availability of one base station chosen at random inside the cluster cell, we first need an expression for the  distribution of the unconditioned size $x$ of a Voronoi cluster.  As there is no exact result known for the size distribution of the Poisson-Voronoi cell, we make use of a two-parameter Gamma function fit of the distribution of the normalized cell size, derived in \cite{ferenc2007},

\footnotesize\begin{eqnarray}\label{eqn:normdens}
f_X(x) = \frac{3.5^{3.5}}{\Gamma[3.5]}x^{2.5}\exp{(-3.5x)}.
\end{eqnarray}\normalsize
Let I be the indicator that a base station chosen at random is located inside a cluster cell. The PDF of the  size conditioned on $I=1$ is derived from $f_X(x)$

\footnotesize\begin{eqnarray}
\nonumber f_{X|I=1}(t) = \frac{f_{X,I=1}(t)}{\bbP[I=1]} = \frac{\bbP[I=1|X=t]f_x(t)}{\bbP[I=1]} \\
\stackrel{(a)}= c t f_X(t) = \frac{3.5^{4.5}}{\Gamma[4.5]}t^{3.5}\exp{(-3.5t)},
\end{eqnarray}\normalsize
where $c$ is a constant such that $\int_{0}^{\infty}{f_{X|I=1}(t)dt} = 1$, and $(a)$ is derived knowing that a randomly chosen base station in $\Pi_\rmb$ is uniformly distributed in the plane of interest and its probability of being inside an area with a given size is proportional to that size. 
The PMF of the number of interferers $N$ inside a cluster cell containing one randomly chosen base station is thus 

\footnotesize\begin{eqnarray}
\nonumber p_{N}(n) &=& \int_{0}^{\infty}{\bbP\left[N = n\; | t\right]f_{X|I=1}(t) \rmd t}\\
\nonumber &\stackrel{(a)}=& \int_{0}^{\infty}{\frac{(\lambda_\rmb t)^{n}e^{-\lambda_\rmb t}}{(n)!} \frac{3.5^{4.5}}{\Gamma[4.5]}t^{3.5}\exp{(-3.5t)}\rmd t}\\
&=& \frac{3.5^{4.5}\Gamma(n+4.5)(\lambda_\rmb/\lambda_\rmc)^n}{\Gamma(4.5)n!(\lambda_\rmb/\lambda_\rmc + 3.5)^{n+4.5}}.
\end{eqnarray}\normalsize

\section{Proof of Result \ref{lem:deltaic}}\label{app:app_4}
For equal-bit-allocation, the expected residual interference is bounded as

\footnotesize\begin{align}
\nonumber &\bbE\bigg\{\sum_{\ell=1}^{N}(1+r_{0,\ell})^{-\alpha}\Gamma\left(\frac{2\Nt-1}{\Nt - 1}\right) 2^{-\lfloor\frac{\rmB_{\mathrm{tot}}}{N+1}\rfloor\frac{1}{\Nt-1}} \bigg\} \\
\nonumber &= \bbE\bigg\{\Nt\Gamma\left(\frac{2\Nt-1}{\Nt - 1}\right)2^{-\lfloor\frac{\rmB_{\mathrm{tot}}}{N+1}\rfloor\frac{1}{\Nt-1}}\sum_{\ell=1}^{N}(1+r_{0,\ell})^{-\alpha}\bigg\}\\
\nonumber &\stackrel{(a)}\leq \bbE\bigg\{\Gamma\left(\frac{2\Nt-1}{\Nt - 1}\right)2^{-\lfloor\frac{\rmB_{\mathrm{tot}}}{N+1}\rfloor\frac{1}{\Nt-1}}N\bbE\bigg\{(1+r_{0,1})^{-\alpha}\bigg\}\bigg\}
\end{align}
\begin{align}
&\stackrel{(b)}= \sum_{n=0}^{\infty}\bigg\{\Gamma\left(\frac{2\Nt-1}{\Nt - 1}\right)n2^{-\lfloor\frac{\rmB_{\mathrm{tot}}}{n+1}\rfloor\frac{1}{\Nt-1}} P_N(n)\bigg\}\bbE\bigg\{(1+r_{0,1})^{-\alpha}\bigg\},
\end{align}\normalsize
where $(a)$ follows by upper bounding the sum of path-loss functions from the interfering cells inside each cluster cell by the path-loss function from the closest interfering base station, $r_{0,1}$. Because the distances $r_{0,\ell}^2$ are 1-D PPP with intensity $\pi\lb$, the random variable $\pi\lambda_\rmb r_{0,1}^2$ has an exponential distribution with parameter 1, $r_{0,1}$ thus has a Rayleigh fading distribution $r_{0,1} \sim \mathrm{Rayleigh}(1/\sqrt{2\pi\lambda_\rmb})$. Therefore

\footnotesize\begin{eqnarray}
\nonumber &\bbE\bigg\{(1+r_{0,1})^{-\alpha}\bigg\} = \\
\nonumber &\int_{0}^{\infty}{\int_{r_0}^{\infty}{(1+x)^{-\alpha}\frac{x}{2\pi\lambda_\rmb}e^{-\frac{x^2}{4\pi\lambda_\rmb}}f_{r_0}(r_0)\rmd x \rmd r_0}}.
\end{eqnarray}\normalsize
$(b)$ follows from expressing the expectation with respect to $N$ by its definition as a function of the PMF of $N$, $P_N(n)$.
To compute $\bbE\bigg\{\log_2\left(\sigma^2 + I_{\mathrm{out}} \right)\bigg\}$, we invoke the Gamma approximation of $I_{\mathrm{out}}$.
The $\Gamma(k,\theta)$ random variable with the same mean and variance as $I_{\mathrm{out}}$ with $|\bg^*_{0,\ell}\bff_{\ell}|^2 \sim \exp(1)$ has the parameters $k$ and $\theta$ given by $k = \frac{(\bbE\{I_{\mathrm{out}}\})^2}{\mbox{var}(x)}$ and $\theta = \frac{\mbox{var}(x)}{\bbE\{I_{\mathrm{out}}\}}$, respectively.
The expected value of $I_{\mathrm{out}} = k \theta$; the logarithm of the interference $I_{\mathrm{out}} = \Gamma(k,\theta)$, with a high interference-to-noise ratio (INR) approximation is given by

\small\begin{eqnarray}
\bbE\bigg\{\log_2(I_{\mathrm{out}})\bigg\}=\log_e(2)\psi(k) + \log_2(\theta).
\end{eqnarray}\normalsize
where  $\psi(k)$ is the digamma function.

\section{Proof of Result \ref{res:DominantInterCluster}}\label{app:app_5}
We consider the terms in the mean loss in rate that correspond to $\rmB_0$. Per spatial realization, given $N$ and $r_{0,\ell}$, the objective function reduces to

\footnotesize\begin{eqnarray}
 &\log_2(e)\Gamma\left(\frac{\Nt}{\Nt-1}\right)2^{-\frac{\rmB_{0}}{\Nt-1}} ~+~\\
\nonumber&\log_2(e) \frac{\Gamma\left(\frac{2\Nt-1}{\Nt-1}\right)}{\bbE\{I_{\mathrm{out}}\} + \invSNR} |\cK|2^{-\frac{\rmB_{\mathrm{tot}} - \rmB_0}{|\cK|(\Nt-1)}}\prod_{\ell\in\cK}{(1 + r_{0,\ell})^{-\frac{\alpha}{|\cK|}})}.
\end{eqnarray}\normalsize
We rewrite the objective function as

\small\begin{eqnarray}\label{eqn:c0}
 \nonumber &2^{-\frac{\rmB_{0}}{\Nt-1}} ~+~ \\
\nonumber &\frac{\Gamma\left(\frac{2\Nt-1}{\Nt-1}\right)}{\bbE\{I_{\mathrm{out}}\} + \invSNR} \nonumber \frac{|\cK|2^{-\frac{\rmB_{\mathrm{tot}}}{|\cK|(\Nt-1)}}\prod_{\ell\in\cK}{(r_{0,\ell}^{-\frac{\alpha}{|\cK|}})}}{\Gamma\left(\frac{\Nt}{\Nt-1}\right)} 2^{\frac{\rmB_0}{|\cK|(\Nt-1)}}\\
  &= 2^{-\frac{\rmB_{0}}{\Nt-1}} ~+~ C_0 2^{\frac{\rmB_0}{|\cK|(\Nt-1)}},
\end{eqnarray}\normalsize
and we invoke the arithmetic mean-geometric mean inequality \small$$2^{-\frac{\rmB_{0}}{\Nt-1}} ~+~ C_0 2^{\frac{\rmB_0}{|\cK|(\Nt-1)}} \leq 2\sqrt{2^{-\frac{\rmB_{0}}{\Nt-1}} C_0 2^{\frac{\rmB_0}{|\cK|(\Nt-1)}}}$$\normalsize to find the minimum of the objective function. We solve for $\rmB_0$ that satisfies the equality \small$2^{-\frac{\rmB_{0}}{\Nt-1}} =  C_0 2^{\frac{\rmB_0}{|\cK|(\Nt-1)}}$\normalsize,
\small$$\rmB_0 = -(\Nt - 1)\log_2(C_0)\frac{|\cK|}{|\cK|+1}.$$\normalsize
Replacing $C_0$ by its value from (\ref{eqn:c0}) yields the result for the low SNR approximation.

\bibliographystyle{IEEEtran}
\bibliography{IEEEabrv,ReferencesSalam}

\begin{thebibliography}{10}
\providecommand{\url}[1]{#1}
\csname url@samestyle\endcsname
\providecommand{\newblock}{\relax}
\providecommand{\bibinfo}[2]{#2}
\providecommand{\BIBentrySTDinterwordspacing}{\spaceskip=0pt\relax}
\providecommand{\BIBentryALTinterwordstretchfactor}{4}
\providecommand{\BIBentryALTinterwordspacing}{\spaceskip=\fontdimen2\font plus
\BIBentryALTinterwordstretchfactor\fontdimen3\font minus
  \fontdimen4\font\relax}
\providecommand{\BIBforeignlanguage}[2]{{%
\expandafter\ifx\csname l@#1\endcsname\relax
\typeout{** WARNING: IEEEtran.bst: No hyphenation pattern has been}%
\typeout{** loaded for the language `#1'. Using the pattern for}%
\typeout{** the default language instead.}%
\else
\language=\csname l@#1\endcsname
\fi
#2}}
\providecommand{\BIBdecl}{\relax}
\BIBdecl

\bibitem{Ges2010}
D.~Gesbert, S.~Hanly, H.~Huang, S.~Shamai, O.~Simeone, and W.~Yu, ``Multi-cell
  {MIMO} cooperative networks: A new look at interference,'' \emph{{IEEE} J.
  Select. Areas Commun.}, vol.~28, no.~9, pp. 1380-- 1408, Dec. 2010.

\bibitem{Papadogiannis2008}
A.~Papadogiannis, D.~Gesbert, and E.~Hardouin, ``A dynamic clustering approach
  in wireless networks with multi-cell cooperative processing,'' in
  \emph{{Proc.} of {IEEE} Int. Conf. on Commun.}, May 2008, pp. 4033--4037.

\bibitem{Lozano2012}
A.~Lozano, J.~G. Andrews, and R.~W. Heath~Jr, ``On the limitations of
  cooperation in wireless networks,'' in \emph{{Proc.} of Information Theory
  and Application Workshop (ITA)}, Feb. 2012.

\bibitem{simeone2009}
O.~Simeone, O.~Somekh, H.~Poor, and S.~Shamai, ``Local base station cooperation
  via finite-capacity links for the uplink of wireless networks,'' \emph{{IEEE}
  Trans. Inform. Theory}, vol.~55, no.~1, pp. 190--204, Jan. 2009.

\bibitem{Zhang2009a}
J.~Zhang, R.~Chen, J.~G. Andrews, A.~Ghosh, and R.~W. Heath~Jr., ``Networked
  {MIMO} with clustered linear precoding,'' \emph{{IEEE} Trans. Wireless
  Commun.}, vol.~8, no.~4, pp. 1910--1921, Apr. 2009.

\bibitem{Barbieri2012}
A.~Barbieri, P.~Gaal, T.~J. Geirhofer, D.~Malladi, Y.~Wei, and F.~Xue,
  ``Coordinated downlink multi-point communications in heterogeneou cellular
  networks,'' in \emph{{Proc. of the Workshop on Information Theory and Its
  Applications}}, 2012.

\bibitem{Shamai2001}
S.~Shamai and B.~M. Zaidel, ``Enhancing the cellular downlink capacity via
  co-processing at the transmitting end,'' in \emph{{Proc.} of {IEEE} Veh.
  Technol. Conf. - Spring}, vol.~3, May 6--9, 2001, pp. 1745--1749.

\bibitem{ekbal2005}
J.~Ekbal and J.~M. Cioffi, ``Distributed transmit beamforming in cellular
  networks - a convex optimization perspective,'' in \emph{{Proc.} of {IEEE}
  Int. Conf. on Commun.}, vol.~4, 2005, pp. 2690--2694.

\bibitem{Jorwiesk2008}
E.~Jorwiesk, E.~G. Larsson, and D.~Danev, ``Complete characterization of the
  pareto boundary for the {MISO} interference channel,'' \emph{{IEEE} Trans.
  Signal Processing}, vol.~56, no.~10, pp. 5292--5296, Oct. 2008.

\bibitem{Ng2008}
B.~L. Ng, J.~S. Evans, S.~V. Hanly, and D.~Aktas, ``Distributed downlink
  beamforming with cooperative base stations,'' \emph{{IEEE} Trans. Inform.
  Theory}, vol.~54, no.~12, pp. 5491--5499, Dec. 2008.

\bibitem{Ng2010}
C.~K. Ng and H.~Huang, ``Linear precoding in cooperative {MIMO} cellular
  networks with limited coordination clusters,'' \emph{{IEEE} J. Select. Areas
  Commun.}, vol.~28, no.~9, pp. 1446 -- 1454, Dec. 2010.

\bibitem{Zhang2010}
J.~Zhang and J.~G. Andrews, ``Adaptive spatial intercell interference
  cancellation in multicell wireless networks,'' \emph{{IEEE} J. Select. Areas
  Commun.}, vol.~28, no.~9, pp. 1455 -- 1468, Dec. 2010.

\bibitem{bhagavatula2011}
R.~Bhagavatula and R.~Heath~Jr, ``Adaptive bit partitioning for multicell
  intercell interference nulling with delayed limited feedback,'' \emph{{IEEE}
  Trans. Signal Processing}, vol.~59, no.~8, pp. 3824--3836, Aug. 2011.

\bibitem{Kaviani2012}
S.~Kaviani, O.~Simeone, W.~Krzymien, and S.~Shamai, ``Linear precoding and
  equalization for network {MIMO} with partial coordination,'' \emph{{IEEE}
  Trans. Veh. Technol.}, vol.~PP, no.~99, pp. 1--1, Feb. 2012.

\bibitem{Foschini2006}
G.~Foschini, K.~Karakayali, and R.~Valenzuela, ``Coordinating multiple antenna
  cellular networks to achieve enormous spectral efficiency,'' \emph{{IEEE
  Proc. Comm.}}, vol. 153, pp. 548--555, Aug. 2006.

\bibitem{Andrews2010}
J.~G. Andrews, F.~Baccelli, and R.~Krishna~Ganti, ``A tractable approach to
  coverage and rate in cellular networks,'' \emph{IEEE Transactions on
  Communications}, vol.~59, no.~11, pp. 3122--3134, Nov. 2011.

\bibitem{CoMP2011}
P.~Marsch and G.~Fettweis, Eds., \emph{Coordinated Multi-Point in Mobile
  Communications. From Theory to Practice}, 1st~ed.\hskip 1em plus 0.5em minus
  0.4em\relax Cambridge University Press, 2011.

\bibitem{baccelli1997}
F.~Baccelli, M.~Klein, M.~Lebourges, and S.~Zuyev, ``Stochastic geometry and
  architecture of communication networks,'' \emph{Journal of Telecommunications
  Systems}, vol.~7, no.~1, pp. 209--227, 1997.

\bibitem{Huang2012j}
\BIBentryALTinterwordspacing
K.~Huang and J.~G. Andrews, ``A closer look at multi-cell cooperation via
  stochastic geometry and large deviations,'' \emph{submitted to {IEEE} Trans.
  Inform. Theory}, Apr. 2012. [Online]. Available:
  \url{http://arxiv.org/abs/1204.3167}
\BIBentrySTDinterwordspacing

\bibitem{Love2008}
D.~J. {Love}, R.~W. {Heath}, Jr., V.~K.~N. {Lau}, D.~{Gesbert}, B.~{Rao}, and
  M.~{Andrews}, ``An overview of limited feedback in wireless communication
  systems,'' \emph{{IEEE} J. Select. Areas Commun.}, vol.~26, no.~8, pp.
  1341--1365, Oct. 2008.

\bibitem{Akoum2011a}
S.~Akoum, M.~Kountouris, and R.~W. Heath~Jr, ``On imperfect {CSI} for the
  downlink of a two-tier network,'' in \emph{{Proc.} of {IEEE} Intl. Symp. on
  Info. Theory}, Jul. 2011, pp. 553--557.

\bibitem{Ozbek2010}
B.~Ozbek and D.~Le~Ruyet, ``Adaptive limited feedback for intercell
  interference cancelation in cooperative downlink multicell networks,'' in
  \emph{{Proc.} of Intnl. Symp. on Wireless Comm. Systems}, Sept. 2010, pp.
  81--85.

\bibitem{Baccelli2009a}
F.~Baccelli and B.~Blaszczyszyn, \emph{Stochastic geometry and wireless
  networks. Volume I: theory}.\hskip 1em plus 0.5em minus 0.4em\relax NOW
  publishers, 2009.

\bibitem{Weber2011}
S.~Weber and J.~G. Andrews, \emph{Transmission Capacity of Wireless Networks},
  Foundations and T.~in~Networking, Eds.\hskip 1em plus 0.5em minus 0.4em\relax
  NOW Publishers, Feb. 2012.

\bibitem{Jindal2011}
N.~Jindal, J.~G. Andrews, and S.~Weber, ``Multi-antenna communication in ad hoc
  networks: achieving {MIMO} gains with {SIMO} transmission,'' \emph{{IEEE}
  Trans. Commun.}, vol.~59, pp. 529--540, Feb. 2011.

\bibitem{Akoum2011}
S.~Akoum, M.~Kountouris, M.~Debbah, and R.~Heath~Jr, ``Spatial interference
  mitigation for multiple input multiple output ad hoc networks: {MISO}
  gains,'' in \emph{{Proc.} of {IEEE} Asilomar Conf. on Signals, Systems, and
  Computers}, Nov. 2011.

\bibitem{Sharif2005}
M.~Sharif and B.~Hassibi, ``On the capacity of {MIMO} broadcast channels with
  partial side information,'' \emph{{IEEE} Trans. Inform. Theory}, vol.~51,
  no.~2, pp. 506--522, 2005.

\bibitem{bhagavatula2010}
R.~Bhagavatula, R.~Heath~Jr, and B.~Rao, ``Limited feedback with joint {CSI}
  quantization for multicell cooperative generalized eigen vector
  beamforming,'' in \emph{{Proc.} of {IEEE} Intl. Conf. on Acoustics Speech and
  Signal Proc.}, Mar. 2010.

\bibitem{ferenc2007}
J.-S. Ferenc and Z.~Neda, ``On the size distribution of {Poisson} {Voronoi}
  cells,'' \emph{Physica A}, vol. 385, pp. 518--526, 2007.

\bibitem{Foss96}
S.~G. Foss and S.~A. Zuyev, ``On a {Voronoi} aggregative process related to a
  bivariate poisson process,'' \emph{{Adv. Appl. Prob.}}, vol.~28, pp.
  965--981, 1996.

\bibitem{Jindal2006}
N.~Jindal, ``{MIMO} broadcast channels with finite rate feedback,''
  \emph{{IEEE} Trans. Inform. Theory}, vol.~52, no.~11, pp. 5045--5060, Nov.
  2006.

\bibitem{Caire2010a}
G.~Caire, N.~Jindal, M.~Kobayashi, and N.~Ravindran, ``Multiuser {MIMO}
  achievable rates with downlink training and channel state feedback,''
  \emph{IEEE Transactions on Information Theory}, vol.~56, no.~6, pp.
  2845--2866, Jun. 2010.

\bibitem{Heath2011}
R.~W. Heath~Jr., T.~Wu, and Y.~H. Kwon, ``Multi-user {MIMO} in distributed
  antenna systems with out-of-cell interference,'' \emph{{IEEE} Trans. Signal
  Processing}, vol.~59, no.~10, pp. 4885--4899, Oct. 2011.

\bibitem{Stoyan1995}
D.~Stoyan, W.~Kendall, and J.~Mecke, \emph{Stochastic Geometry and Its
  Applications}.\hskip 1em plus 0.5em minus 0.4em\relax John Wiley and Sons,
  Ltd., 1995.

\bibitem{Calka2002}
P.~Calka, ``The distributions of the smallest disks containing the
  {Poisson-Voronoi} typical cell and the {Crofton} cell in the plane,''
  \emph{Advances in Applied Probability}, vol.~34, no.~4, pp. 702--717, Dec.
  2002.

\bibitem{baccelli2009}
F.~Baccelli, B.~Blaszczyszyn, and P.~Muhlethater, ``Stochastic analysis of
  spatial and opportunistic {Aloha},'' \emph{{IEEE} J. Select. Areas Commun.},
  vol.~27, no.~7, pp. 1105--1119, sept. 2009.

\end{thebibliography}

\end{document}